\documentclass[10pt,twocolumn,letterpaper]{article}

\usepackage{cvpr}              %

\usepackage[accsupp]{axessibility}

\usepackage{pifont}
\newcommand{\xmark}{\ding{55}}

\newbool{authorcomments}
\booltrue{authorcomments} %
\ifbool{authorcomments}%
{
\newcommand{\red}[1]{{\color{red}#1}}
\newcommand{\todo}[1]{{\color{red}#1}}
\newcommand{\TODO}[1]{\textbf{\color{red}[TODO: #1]}}
\newcommand{\zg}[1]{{{\color{blue}[Zan: #1]}}}
\newcommand{\nm}[1]{{{\color{orange}[Nico: #1]}}}
\newcommand{\qw}[1]{{{\color{teal}[Qi: #1]}}}
\newcommand{\jm}[1]{{{\color{purple}[Janick: #1]}}}
}
{
\newcommand{\red}[1]{{}}
\newcommand{\todo}[1]{{}}
\newcommand{\TODO}[1]{{}}
\newcommand{\zg}[1]{{}}
\newcommand{\nm}[1]{{}}
\newcommand{\qw}[1]{{}}
\newcommand{\jm}[1]{{}}
}

\usepackage[normalem]{ulem}

\newcommand{\revAdded}[1]{{#1}}
\newcommand{\revRemoved}[1]{{}}
\newcommand{\revReplaced}[2]{#2}

\newcommand{\RayOrigin}{\bm{o}}
\newcommand{\RayDirection}{\bm{d}}
\newcommand{\ParticleCenter}{\bm{\mu}}
\newcommand{\ParticleScale}{\bm{s}}
\newcommand{\ParticleRotation}{\bm{q}}
\newcommand{\ParticleRadiance}{\phi}
\newcommand{\ParticleOpacity}{\sigma}
\newcommand{\ParticleHarmonics}{\bm{\beta}}
\newcommand{\ParticleResponse}{\rho}

\newcommand{\TMax}{\tau_\textrm{max}}

\definecolor{gold}{RGB}{221, 196, 65}
\definecolor{silver}{RGB}{215, 215, 215}
\definecolor{bronze}{RGB}{126, 66, 5}
\definecolor{green_zoomin}{RGB}{42, 137, 0}
\definecolor{nsteps_blue}{rgb}{0.03137254901960784, 0.18823529411764706, 0.4196078431372549}

\definecolor{table_red}{rgb}{1, 0.7, 0.7}
\definecolor{table_orange}{rgb}{1,0.85, 0.7}
\definecolor{table_yellow}{rgb}{1,1, 0.8}

\newcommand{\ourmethod}{\textsc{3DGUT}\xspace}
\usepackage{colortbl, xcolor}
\newcommand{\parahead}[1]{\vspace{1mm}\noindent\textbf{{#1}.}\ }

\usepackage{algorithmicx}
\usepackage{algpseudocode}
\algtext*{EndProcedure}
\algtext*{EndIf}
\algtext*{EndFor}
\newcommand{\InputConditions}[1]{\Require \parbox[t]{\dimexpr\linewidth-\algorithmicindent}{#1\strut}}
\newcommand{\OutputConditions}[1]{\Ensure \parbox[t]{\dimexpr\linewidth-\algorithmicindent}{#1\strut}}

\newcommand{\Proc}[1]{\mbox{\textsc{#1}}}

\definecolor{commentblue}{rgb}{0.407,0.663,.800}
\renewcommand{\Comment}[1]{\hfill\textcolor{commentblue}{\(\triangleright\)\textit{#1}}}

\newcounter{algo}
\newenvironment{algo}[1]
{ \refstepcounter{algo}\noindent\rule{\columnwidth}{1.25pt}\vspace{-.2\baselineskip} \\ \textbf{Algorithm~\thealgo} #1\vspace{-.55\baselineskip} \\ \noindent\rule{\columnwidth}{.5pt}\vspace{-1.2\baselineskip} }
{ \vspace{-.8\baselineskip}\noindent\rule{\columnwidth}{.5pt}\vspace{-\baselineskip} }

\algnewcommand{\LeftComment}[1]{\textcolor{commentblue}{\(\triangleright\)\textit{#1}}}

\newcommand\blfootnote[1]{%
  \begingroup
  \renewcommand\thefootnote{}\footnote{#1}%
  \addtocounter{footnote}{-1}%
  \endgroup
}

\usepackage{amsmath,amsfonts,bm}

\def\1{\bm{1}}

\def\vc{{\bm{c}}}

\def\vx{{\bm{x}}}

\DeclareMathAlphabet{\mathsfit}{\encodingdefault}{\sfdefault}{m}{sl}
\SetMathAlphabet{\mathsfit}{bold}{\encodingdefault}{\sfdefault}{bx}{n}

\graphicspath{{fig/}%
{fig/compressed}%
}

\definecolor{cvprblue}{rgb}{0.21,0.49,0.74}
\usepackage[pagebackref,breaklinks,colorlinks,allcolors=cvprblue]{hyperref}
\usepackage{multirow}

\def\paperID{*****} %
\def\confName{-}
\def\confYear{2025}

\title{3DGUT: Enabling Distorted Cameras and Secondary Rays in Gaussian Splatting \vspace{-11mm}}
\author{
Qi Wu$^{1}$\footnotemark[1], Janick Martinez Esturo$^{1}$\footnotemark[1], Ashkan Mirzaei$^{1,2}$, Nicolas Moenne-Loccoz$^{1}$, Zan Gojcic$^{1}$ \\
\small
$^{1}$NVIDIA, $^{2}$University of Toronto \\
\small\textbf{\url{https://research.nvidia.com/labs/toronto-ai/3DGUT}}\\
}

\begin{document}

\twocolumn[{
\maketitle
\vspace{-11mm}
\renewcommand\twocolumn[1][]{#1}
\begin{center}
    \centering
    \includegraphics[width=\linewidth]{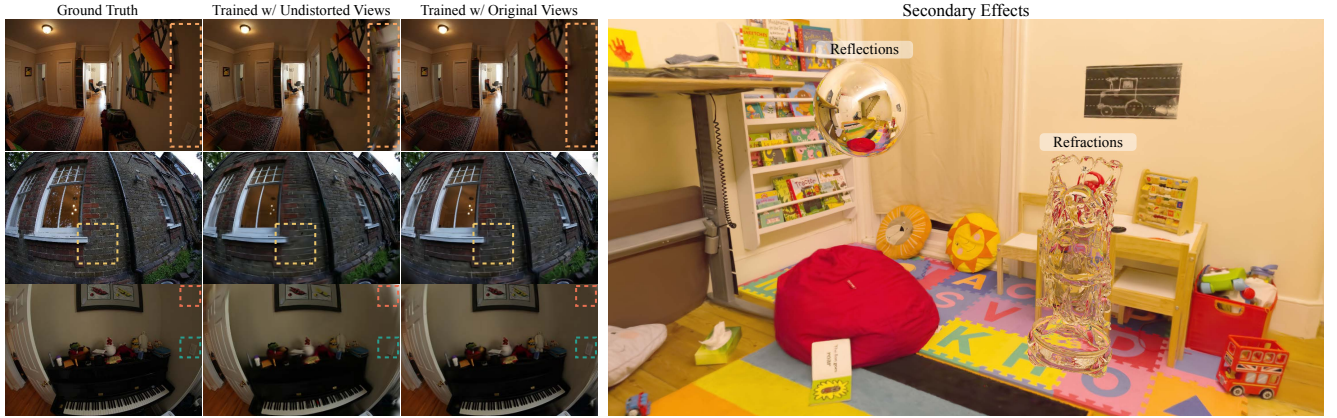}
    \vspace{-6mm}
    \captionof{figure}{We extend 3D Gaussian Splatting (3DGS) to support nonlinear camera projections and secondary rays for simulating phenomena such as reflections and refractions. By replacing EWA splatting rasterization with the Unscented Transform, our approach retains real-time efficiency while accommodating complex camera effects like rolling shutter. (Left) A comparison of our model trained on undistorted views vs. the original distorted fisheye views, showing that training on the full set of pixels improves visual quality. (Right) Two synthetic objects, a reflective sphere and a refractive statue, inserted into a scene reconstructed with our model.
    }
    \label{fig:teaser}
\end{center}
}]
\blfootnote{\hspace{-1em} $^*$ denotes equal contribution.}

\begin{abstract}
3D Gaussian Splatting (3DGS) enables efficient reconstruction and high-fidelity real-time rendering of complex scenes on consumer hardware.  However, due to its rasterization-based formulation, 3DGS is constrained to ideal pinhole cameras and lacks support for secondary lighting effects. Recent methods address these limitations by tracing the particles instead, but, this comes at the cost of significantly slower rendering. In this work, we propose 3D Gaussian Unscented Transform (3DGUT), replacing the EWA splatting formulation with the Unscented Transform that approximates the particles through sigma points, which can be projected exactly under any nonlinear projection function. This modification enables trivial support of distorted cameras with time dependent effects such as rolling shutter, while retaining the efficiency of rasterization. Additionally, we align our rendering formulation with that of tracing-based methods, enabling secondary ray tracing required to represent phenomena such as reflections and refraction within the same 3D representation.  \revAdded{The source code is available at: }{\small\textbf{\url{https://github.com/nv-tlabs/3dgrut}}}.
\end{abstract}
    
\section{Introduction}
\label{sec:intro}
Multiview 3D reconstruction and novel view synthesis is a classical problem in computer vision, for which several scene representations have been proposed in recent years, including points~\cite{KPLD21, ruckert2022adop}, surfaces~\cite{Riegler2020FVS, chen2022mobilenerf, yariv2023bakedsdf}, and volumetric fields~\cite{mildenhall2020nerf, wang2021neus, yariv2021volume, mueller2022instant}. Most recently, driven by 3D Gaussian Splatting~\cite{kerbl3Dgaussians} (3DGS), volumetric particle-based representations have gained significant popularity due to their high visual fidelity and fast rendering speeds. The core idea of 3DGS is to model scenes as an unstructured collection of fuzzy 3D Gaussian particles, each defined by its location, scale, rotation, opacity, and appearance. These particles can be rendered differentiably in real time via rasterization, allowing their parameters to be optimized through a re-rendering loss function. 

High frame-rates of 3DGS, especially compared to volumetric ray marching methods, can be largely accredited to the efficient rasterization of particles. However, this reliance on rasterization also imposes some inherent limitations. The EWA splatting formulation \cite{zwicker2002ewa} does not support highly-distorted cameras with complex time dependent effects such as rolling shutter. Additionally, rasterization cannot simulate secondary rays required for representing phenomena like reflection, refraction, and shadows. 

Instead of rasterization, recent works have proposed to render the volumetric particles using ray tracing~\cite{3dgrt2024, Condor2024Gaussians, mai2024ever}. While this mitigates the shortcomings of rasterization, it does so at the expense of significantly reduced rendering speed, even when the tracing formulation is heavily optimized for semi-transparent particles~\cite{3dgrt2024}. In this work, we instead aim to overcome the above limitations of 3DGS while remaining in the realm of rasterization, thereby maintaining the high-rendering rates. To this end, we seek answer to the following two questions: 

\emph{What makes 3DGS ill-suited to represent distorted cameras and rolling shutter?} 
To project 3D Gaussian particles onto the camera image plane, 3DGS relies on an EWA splatting formulation that requires computing the Jacobian of the non-linear projection function. 
This leads to approximation errors, even for perfect pinhole cameras, and the errors become progressively worse with increasing distortion~\cite{huang2024erroranalysis3dgaussian}. Moreover, it is unclear how to even represent time-dependent effect such as rolling-shutter within the EWA splatting formulation. 

Instead of approximating the non-linear projection function, we draw inspiration from the classical literature of Unscented Kalman Filter~\cite{julier1997UKF} and approximate the 3D Gaussian particles using a set of carefully selected sigma points. These sigma points can be projected exactly onto the camera image plane by applying an arbitrarily complex projection function to each point, after which a 2D Gaussian can be re-estimated from them in form of a Unscented Transform (UT)~\cite{gustafsson2012UT}. Apart from a better approximation quality, UT is derivative-free and completely avoids the need to derive the Jacobians for different camera models (\cref{fig:teaser} left). Moreover, complex effects such as rolling shutter distortions can directly be represented by transforming each sigma point with a different extrinsic matrix. 

\emph{Can we align the rasterization rendering formulation with the one of ray-tracing?} The rendering formulations mainly differ in terms of: (i) determining which particles contribute to which pixels, (ii) the order in which the particles are intersected, (iii) how the particles are evaluated.
To align the representations we therefore follow 3DGRT~\cite{3dgrt2024} and evaluate the Gaussian particle response in 3D, while sorting them in order similar to ~\citet{radl2024stopthepop}. While small differences persist, this provides us with a representation that can be both rasterized and ray-traced, enabling secondary-rays required to simulate phenomena like refraction and reflection (\cref{fig:teaser} right).

In summary, we propose 3D Gaussian Unscented Transform (3DGUT), where our main contributions are:
\begin{itemize}[topsep=1pt, leftmargin=10pt]
\item We derive a rasterization formulation that approximates the 3D Gaussian particles instead of \revRemoved{approximating }the non-linear projection function. This simple change enables us to extend 3DGS to arbitrary camera models and to support complex time dependent effects such as rolling shutter.
\item We align the rendering formulation with 3DGRT, which allows us to render the same representation with rasterization and ray-tracing, supporting phenomena such as refraction and reflections.
\end{itemize}

On multiple datasets, we demonstrate that our formulation leads to comparable rendering rates and image fidelity to 3DGS, while offering greater flexibility and outperforming dedicated methods on datasets with distorted cameras.

\vspace{-3mm}
\section{Related Work}
\label{sec:formatting}
\paragraph{Neural Radiance Fields}
Neural Radiance Fields (NeRFs)~\cite{mildenhall2020nerf} have transformed the field of novel view synthesis, by modeling scenes as emissive volume encoded within coordinate-based neural network. These networks can be queried at any spatial location to return the volume density and view-dependent radiance. Novel views are synthesized by sampling the network along camera rays and accumulating radiance through volumetric rendering. While the original formulation~\cite{mildenhall2020nerf} utilized a large, global multi-layer perceptron (MLP), subsequent work has explored more efficient scene representations, including voxel grids~\cite{liu2020neural, yu_and_fridovichkeil2021plenoxels, SunSC22}, triplanes~\cite{Chan2022}, low-rank tensors~\cite{Chen2022ECCV}, and hash tables~\cite{mueller2022instant}. Despite these advances, even highly optimized NeRF implementations~\cite{mueller2022instant} still struggle to achieve real-time inference rates due to the computational cost of ray marching. 

To accelerate inference, several efforts have focused on converting the radiance fields into more efficient representations, such as meshes~\cite{chen2022mobilenerf, yariv2023bakedsdf}, hybrid surface-volume representations~\cite{wan2023learning, wang2023adaptive, turki2024hybridnerf, sharma2023volumetric}, and sparse volumes~\cite{garbin2021fastnerf, Reiser2023SIGGRAPH, duckworth2023smerf}. However, these approaches generally require a cumbersome two-step pipeline: first training a conventional NeRF model and then baking it into a more performant representation, which further increases the training time and complexity. 

\vspace{-3mm}
\paragraph{Volumetric Particle Representations}
Differentiable rendering via alpha compositing has also been explored in combination with volumetric particles, such as spheres~\cite{lassner2021pulsar}. More recently, 3D Gaussian Splatting~\cite{kerbl3Dgaussians} replaced spheres with fuzzy anisotropic 3D Gaussians. Instead of ray marching, these explicit volumetric particles can be rendered through highly efficient rasterization, achieving competitive results in terms of quality and efficiency. Due to its simplicity and flexibility, 3DGS has inspired numerous follow-up works focusing on improving memory efficiency~\cite{Lee_2024_CVPR, mallick2024taming3dgshighqualityradiance, scaffoldgs}, developing better densification and pruning heuristics~\cite{ye2024absgs, kheradmand20243DGSMCMC}, enhancing surface representation~\cite{guedon2023sugar, guedon2024frosting}, and scaling up to large scenes~\cite{liu2024citygaussian, lin2024vastgaussian, hierarchicalgaussians24}. However, while rasterization is very efficient, it also introduces trade-offs, such as being limited to perfect pinhole cameras. Prior work has attempted to work around these limitations and support complex camera models such as fisheye cameras~\cite{liao2024fisheye} or rolling shutter~\cite{seiskari2024gaussian}. But these works still require dedicated formulation for each camera type and exhibit quality degradation with increased complexity and distortion of the camera models~\cite{huang2024erroranalysis3dgaussian}.

In response, recent works have explored replacing rasterization entirely and instead rendering the 3D Gaussians using ray tracing~\cite{3dgrt2024, Condor2024Gaussians, mai2024ever}. Ray tracing inherently supports complex camera models and enables secondary effects like shadows, refraction, and reflections through secondary rays. However, this comes with a substantial decrease in rendering efficiency: even the most optimized ray-tracing methods are still 3-4 times slower than rasterization~\cite{3dgrt2024}.

In this work, we instead propose a generalized approach for efficiently handling complex camera models within the rasterization framework, thereby preserving the computational efficiency. Additionally we unify our rendering formulation with the one of ray-tracing, enabling a hybrid rendering technique within the same representation.

\vspace{-3mm}
\paragraph{Unscented Transform}
Computing the statistics of a random variable that has undergone a transformation is one of the fundamental tasks in the fields of estimation and optimization. When the transformation is non-linear, however, no closed form solution exists, so several approximations have been proposed
. The simplest and perhaps most widely used approach is to linearize the non-linear transformation using the first order Taylor approximation. However, the local linearity assumption is often violated, and derivation of the Jacobian matrix is non-trivial and error prone. The Unscented Transform (UT)~\cite{julier1995new, julier1997UKF} was proposed to address these limitations. The key idea of UT is to approximate the distribution of the random variable using a set of Sigma points that can be transformed exactly, after which they can be used to re-estimate the statistics of the random variable in the target domain. Originally, UT was devised for filtering-based state estimation~\cite{julier1997UKF, wan2000unscented}, but it has since found applications in computer vision~\cite{janjoš2023unscentedautoencoder, barron2023zipnerf}. Notably, UT has even been explored in the context of novel-view synthesis~\cite{barron2023zipnerf}, where it was used to estimate the ray frustum from samples that match its first and second moments. 

\vspace{-2mm}
\section{Preliminaries}
\begin{figure}
    \centering
    \includegraphics[width=\linewidth]{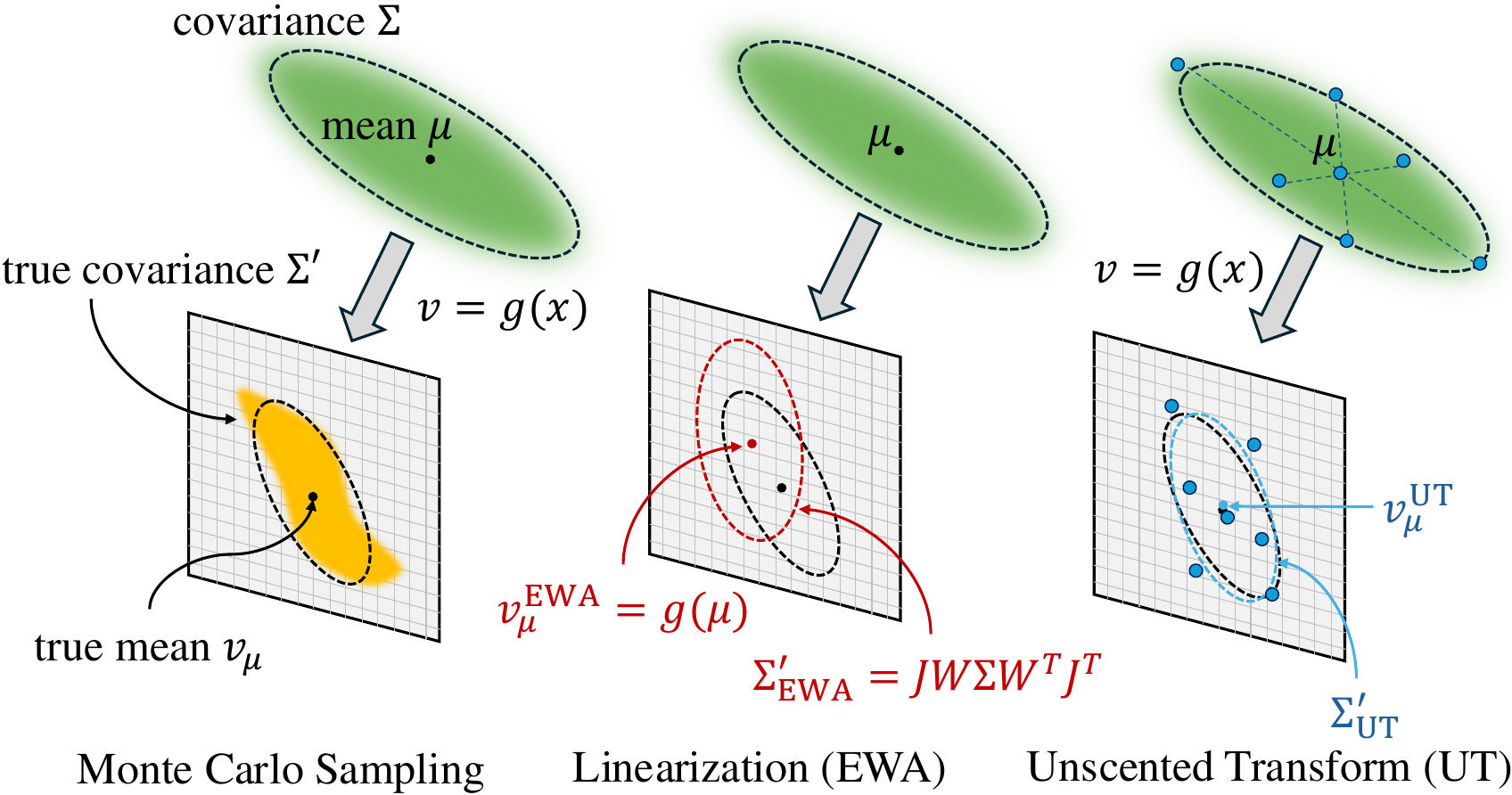}
    \vspace{-6mm}
    \caption{When projecting a Gaussian particle from 3D space onto the camera image plane, Monte Carlo sampling (\emph{left}) provides the most accurate estimate but is costly to compute. EWA Splatting formulation used in~\cite{kerbl3Dgaussians} approximates the projection function via linearization, which requires a dedicated Jacobian $J$ for each camera model and leads to approximation errors with increasing distortion. Unscented Transform instead approximates the particle with Sigma points than can be projected exactly and from which the 2D conic can then be estimated.}
    \label{fig:UT}
    \vspace{-4mm}
\end{figure}

\label{sec:preliminaries}
We provide a short review of 3D Gaussian parametrization, volumetric particle rendering, and EWA splatting.

\vspace{-3mm}
\paragraph{3D Gaussian Splatting Representation:}
\citet{kerbl3Dgaussians} represent scenes using an unordered set of 3D Gaussian particles whose response function $\ParticleResponse:\mathbb{R}^3\rightarrow\mathbb{R}$ is defined as
\begin{equation}
    \label{eq:gaussian_kernel}
    \ParticleResponse(\bm{x}) = \exp(-\frac{1}{2}{(\bm{x}-\ParticleCenter)^T\bm{\Sigma^{-1}}(\bm{x}-\ParticleCenter)}),
\end{equation}
where $\ParticleCenter \in \mathbb{R}^3$ denotes the particle's position and $\bm{\Sigma} \in \mathbb{R}^{3 \times 3}$ its covariance matrix.
To ensure that $\bm{\Sigma}$ remains positive semi-definite during gradient-based optimization, it is decomposed into a rotation matrix $\bm{R} \in \mathrm{SO(3)}$ and a scaling matrix $\bm{S} \in \mathbb{R}^{3 \times 3}$, such that
\begin{equation}
    \label{eq:gaussian_covariance}
    \bm{\Sigma}=\bm{R}\bm{S}\bm{S}^{T}\bm{R}^{T}
\end{equation}
In practice, both $\bm{R}$ and $\bm{S}$ are stored as vectors---a quaternion $\ParticleRotation \in \mathbb{R}^4$ for the rotation and a vector $\ParticleScale \in \mathbb{R}^3$ for the scaling. Each particle is also associated with an opacity coefficient, $\ParticleOpacity \in \mathbb{R}$, and a view-dependent parametric radiance function $\ParticleRadiance_{\ParticleHarmonics}(\RayDirection): \mathbb{R}^3 \rightarrow \mathbb{R}^3$, with $\RayDirection$ the incident ray direction, which is in practice represented using spherical harmonics functions of order $m=3$.

\vspace{-3mm}
\paragraph{Determining the Particle Response:} Within the 3DGS rasterization framework, the 3D particles first need to be projected to the camera image plane in order to determine their contributions to the individual pixels. To this end, 3DGS follows~\cite{zwicker2002ewa} and computes a covariance matrix $\bm{\Sigma'} \in \mathbb{R}^{2 \times 2}$ for a projected Gaussian in image coordinates via first-order approximation as 
\begin{equation}
\label{eq:ewa_splatting}
\bm{\Sigma'}=\bm{J}_{[:2,:3]}\bm{W}\bm{\Sigma}\bm{W}^{T}\bm{J}_{[:2,:3]}^{T}
\end{equation}
where $\bm{W} \in \mathrm{SE(3)}$ transforms the particle from the world to the camera coordinate system, and $\bm{J} \in \mathbb{R}^{3 \times 3}$ denotes the Jacobian matrix of the affine approximation of the projective transformation, which is obtained by considering the linear terms of its Taylor expansion. The Gaussian response of a particle $i$ for a position $\bm{x} \in \mathbb{R}^{3}$ can then be computed in 2D from its projection on the image plane $\bm{v}_{\bm{x}} \in \mathbb{R}^{2}$ as
\begin{equation}
    \label{eq:gaussian_kernel_2d}
    \ParticleResponse_{i}(\bm{x}) = \exp(
        -\frac{1}{2} 
        (\bm{v}_{\bm{x}}-\bm{v}_{\ParticleCenter_{i}})^{T} 
        \bm{{\Sigma'}_{i}^{-1}} 
        (\bm{v}_{\bm{x}}-\bm{v}_{\ParticleCenter_{i}}) 
    )
\end{equation}
where $\bm{v}_{\ParticleCenter_{i}} \in \mathbb{R}^2$ denotes the projected mean of the particle.

\paragraph{Volumetric Particle Rendering:}
The color $\vc \in \mathbb{R}^3$ of a camera ray $\bm{r}(\tau) = \RayOrigin + \tau\RayDirection$ with origin $\RayOrigin \in \mathbb{R}^3$ and direction $\RayDirection \in \mathbb{R}^3$ can be rendered from the above volumetric particle representation using numerical integration
\begin{equation}
\label{eq:QuadratureVolumeRendering}
\vc(\RayOrigin,\RayDirection) = \sum_{i=1}^N \vc_i(\RayDirection) \alpha_i \prod_{j=1}^{i-1} 1 -\alpha_j,
\end{equation}
where $N$ denotes the number of particles that contribute to the given ray and opacity $\alpha_i \in \mathbb{R}$ is defined as $\alpha_i =\sigma_i\ParticleResponse_i(\RayOrigin + \tau\RayDirection)$ for any $\tau \in \mathbb{R}^{+}$.

\section{Method}
\label{sec:method}
Our aim is to extend 3DGS~\cite{kerbl3Dgaussians} and 3DGRT~\cite{3dgrt2024} methods by developing a formulation that:

\begin{itemize} 
\item accommodates highly distorted cameras and time-dependent camera effects, such as rolling shutter,
\item unifies the rendering formulation to allow the same reconstructions to be rendered using either splatting or tracing, enabling hybrid rendering with traced secondary rays,
\end{itemize}

\noindent
all while preserving the efficiency of rasterization. We begin by detailing our approach to bypass the linearization steps of 3DGS~\cite{kerbl3Dgaussians} in \cref{sec:unscented_tranform}, followed by an approach to evaluate the particles in order and directly in 3D (\cref{sec:evaluating_particle_response}). The former enables support for complex camera models, while the latter aligns the rendering formulation with 3DGRT~\cite{3dgrt2024}.

\subsection{Unscented Transform}
\label{sec:unscented_tranform}
As illustrated in \cref{fig:UT}, the EWA splatting formulation used in 3DGS for projecting 3D Gaussian particles onto the camera image plane relies on the linearization of the affine approximation of the projective transform (\cref{eq:ewa_splatting}). This approach, however, has several notable limitations: (i) it neglects higher-order terms in the Taylor expansion, leading to projection errors even with perfect pinhole cameras~\cite{huang2024erroranalysis3dgaussian}, and these errors increase with camera distortion; (ii) it requires deriving a new Jacobian for each specific camera model (e.g., the equidistant fisheye model in~\cite{liao2024fisheye}), which is cumbersome and error prone; (iii) it necessitates representing the projection as a single function, which is particularly challenging when accounting for time-dependent effects such as rolling shutter.

To overcome these limitations, we \revReplaced{build on}{leverage} the idea\revRemoved{s} of the Unscented Transform (UT) and propose to instead approximate the volumetric \revAdded{$N$-dimensional} particle using a set of carefully selected Sigma points. \revAdded{Generally, $2N +1$ points are required to match at least the first three moments of the target distribution}. \revReplaced{Specifically, consider}{Consider} the 3D Gaussian scene representation described in \cref{sec:preliminaries}, where particles are characterized by their position $\ParticleCenter$ and covariance matrix $\bm{\Sigma}$\revReplaced{. The}{, the} Sigma points $\mathcal{X} = \{ \vx_i\}_{i=0}^6$ are then defined as

\begin{align}
    \label{eq:sigma_points}
    \vx_i &= 
    \begin{cases}
        \ParticleCenter & \text{for } i = 0 \\[3pt]
        \ParticleCenter + \sqrt{(3 + \lambda) \bm{\Sigma}}_{[i]} & \text{for } i = 1, 2, 3 \\[3pt]
        \ParticleCenter - \sqrt{(3 + \lambda) \bm{\Sigma}}_{[i-3]} & \text{for } i = 4, 5, 6
    \end{cases}
\end{align}

using the available factorization \cref{eq:gaussian_covariance} of the covariance to read of the matrix square-root.

Their corresponding weights $\mathcal{W} = \{ w_i\}_{i=0}^6$ are given as
\begin{align}
    \label{eq:weights_1}
    w_i^\mu &= 
    \begin{cases}
    \frac{\lambda}{3 + \lambda} & \quad \text{for } i = 0 \\[3pt]
    \frac{1}{2(3 + \lambda)} & \quad \text{for } i = 1, \dots, 6
    \end{cases} \\[10pt]
    w_i^\Sigma &= 
    \label{eq:weights_2}
    \begin{cases}
    \frac{\lambda}{3 + \lambda} + (1 - \alpha^2 + \beta) & \quad \text{for } i = 0 \\[3pt]
    \frac{1}{2(3 + \lambda)} & \quad \text{for } i = 1, \dots, 6
    \end{cases}
\end{align}
where $\lambda = \alpha^2 (3 + \kappa) - 3$, $\alpha$ is a hyperparameter that controls the spread of the points around the mean, $\kappa$ is a scaling parameter typically set to 0, and $\beta$ is used to incorporate prior knowledge about the distribution ~\cite{wan2000unscented}.

Each Sigma point can then be independently projected onto the camera image plane using the non-linear projection function $\bm{v}_{x_i} = g(\vx_i)$. The 2D conic can subsequently be approximated as the weighted posterior sample mean and covariance matrix of the Gaussian:
\begin{align}
    \label{eq:2d_conic_1}
    \bm{v}_{\ParticleCenter} & = \sum_{i=0}^6 w_i^{\mu}\bm{v}_{x_i} \\
    \label{eq:2d_conic_2}
    \bm{\Sigma'} & = \sum_{i=0}^6 w_i^{\Sigma}(\bm{v}_{x_i} - \bm{v}_{\ParticleCenter})(\bm{v}_{x_i} - \bm{v}_{\ParticleCenter})^{\text{T}}
\end{align}
 
With the 2D conic computed, we can apply the same tiling and culling procedures as proposed by~\cite{kerbl3Dgaussians, radl2024stopthepop} to determine which particles influence which pixels. As described in the following section, our particle response evaluation does not depend on the 2D conic. Instead, UT only acts as an acceleration structure to efficiently determine the particles that contribute to each pixel thus avoiding the need for computing the backward pass through the non-linear projection function.

\subsection{Evaluating Particle Response}
\label{sec:evaluating_particle_response}
\begin{figure}
    \centering
    \includegraphics[width=\linewidth]{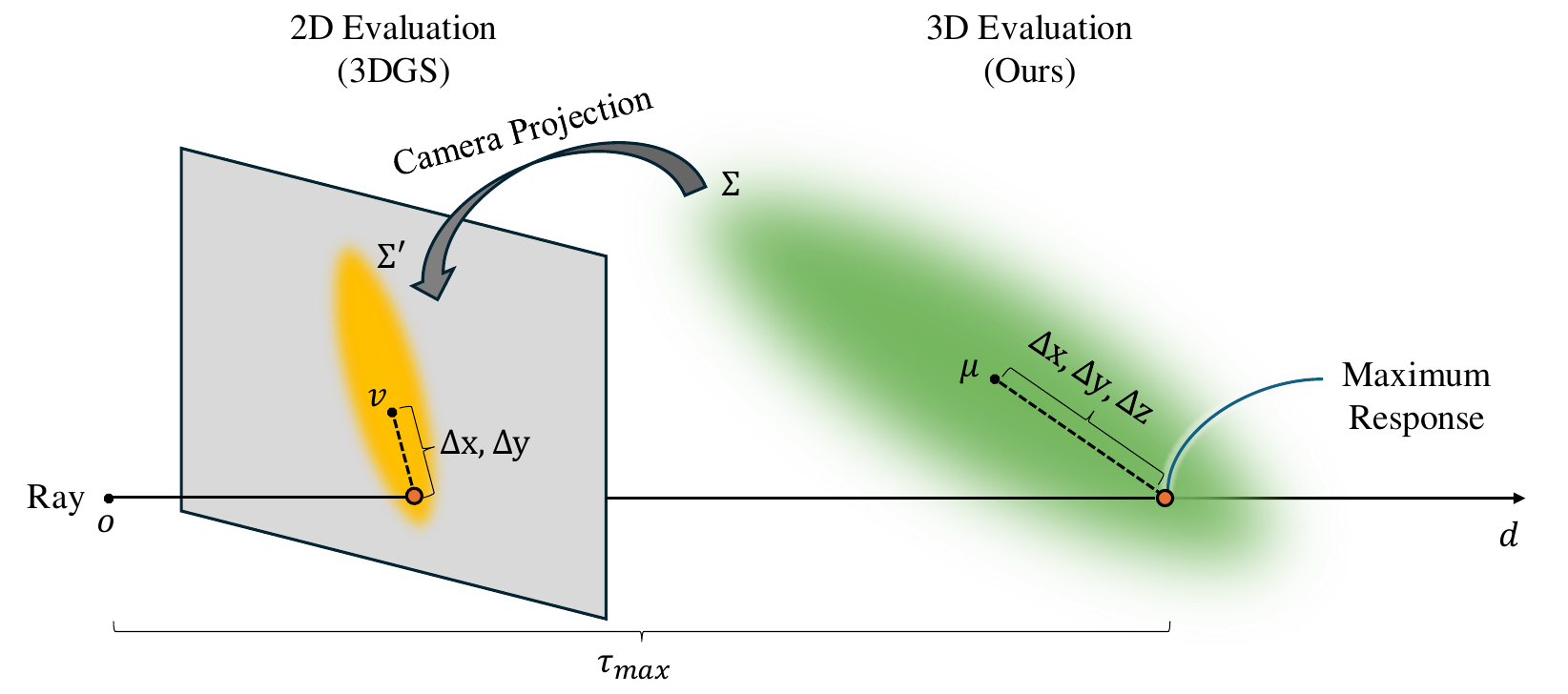}
    \vspace{-7mm}
    \caption{For a given ray, 3DGS~\cite{kerbl3Dgaussians} evaluates the response of the Gaussian particle in 2D after the projection onto the camera image plane. This requires backpropagation through the (approximated) projection function. Instead, we follow \cite{3dgrt2024} and evaluate particles in 3D at the point of the maximum response along the ray.}
    \label{fig:particle_response}
    \vspace{-4mm}
\end{figure}

Once the Gaussian particles contributing to each pixel have been identified, we need to determine how to evaluate their response. Following 3DGRT~\cite{3dgrt2024}, we evaluate particles directly in 3D by using a single sample located at the point of maximum particle response along a given ray.

A comparison between 3DGS's 2D conic response evaluation method  and our 3D response evaluation method is provided in \cref{fig:particle_response}.
Specifically, we compute the distance $\TMax=\textrm{argmax}_{\tau}{\ParticleResponse(\RayOrigin+\tau \RayDirection)}$, which maximizes the particle response along the ray $\bm{r}(\tau)$, as 
\begin{equation}
  \label{eq:GaussianMaxResponse}
  \TMax = \frac{(\ParticleCenter - \RayOrigin)^T\bm{\Sigma}^{-1}\RayDirection}{\RayDirection^T\bm{\Sigma}^{-1}\RayDirection} = \frac{-\RayOrigin_g^T\RayDirection_g}{\RayDirection_g^T \RayDirection_g}
\end{equation}
where $\RayOrigin_g=\bm{S}^{-1}\bm{R}^T(\RayOrigin-\ParticleCenter)$ and $\RayDirection_g=\bm{S}^{-1}\bm{R}^T\RayDirection$.

Unlike 3DGS, which performs particle evaluations in 2D, our approach avoids propagating gradients through the projection function, thereby avoiding the approximations and mitigating potential numerical instabilities. \revAdded{Due to limited space, we provide the derivation of the numerically stable backward pass in the Supplementary Material \cref{sec:derivative}.}

\subsection{Sorting Particles}
\label{sec:unification}
The proposed volumetric rendering formulation, i.e. both the rendering equation \cref{eq:QuadratureVolumeRendering} and the particle evaluation \cref{eq:GaussianMaxResponse}, is equivalent to the one used in 3DGRT. 
However, while 3DGRT is able to collect the hit particles in their exact $\TMax$ order along the ray thanks to a dedicated acceleration structure \cite{parker2010optix}, 3DGS sorts them globally for each tile.
In order to get a better approximation of the $\TMax$ order we propose to use the multi-layers alpha blending approximation (MLAB) \cite{Salvi2014MultilayerAB} following~\cite{radl2024stopthepop}.\footnote{StopThePop~\cite{radl2024stopthepop} denotes \revRemoved{the} MLAB \revRemoved{formulation} as \revAdded{the} $k$-buffer approach.} It consists in storing the per-ray $k$-farthest hit particles (typically using $k$ = 16) in a buffer. The closest hits which cannot be stored in the buffer are incrementally alpha-blended until the transmittance of the blended part vanishes. 

As an alternative, the hybrid transparency (HT) blending strategy \cite{mauleHybrid2013} has been recently used for splatting Gaussian particles \cite{hahlbohm2024efficientperspectivecorrect3dgaussian}.
Instead of storing the $k$-farthest hit particles and incrementally blending the closest hits, HT stores the $k$-closest and incrementally blends the farthest hits. This permits to recover the exact $k$-closest hit particles, but requires to go through all particles, which may be prohibitively slow without dedicated optimizations and heuristics.  

\subsection{Implementation and Training}
\label{sec:training}
We build on the work of \cite{kerbl3Dgaussians, 3dgrt2024} and implemented our method in PyTorch, using custom CUDA kernels for the compute-intensive parts. Additionally, we employ advanced culling strategies proposed by \citet{radl2024stopthepop}. Unless otherwise specified, we adopt all parameters from 3DGS~\cite{kerbl3Dgaussians} to ensure a fair comparison and keep them consistent across all evaluations.

Similar to \cite{3dgrt2024} we don't have access to 2D screen space gradients, so we follow 3DGRT~\cite{3dgrt2024} and replace them with the 3D positional gradients divided by half of the distance to the camera and perform densification and pruning every 300 iterations. For the UT, we set $\alpha=1.0$, $\beta=2.0$ and $\kappa=0.0$ in all evaluations. We train our model for 30k iterations using the weighted sum of the L2-loss $\mathcal{L}_2$ and the perceptual loss $\mathcal{L_\text{SSIM}}$ sucht that $\mathcal{L} = \mathcal{L}_2 + 0.2 \mathcal{L_\text{SSIM}}$.

\begin{figure*}
    \centering
    \includegraphics[width=0.98\linewidth]{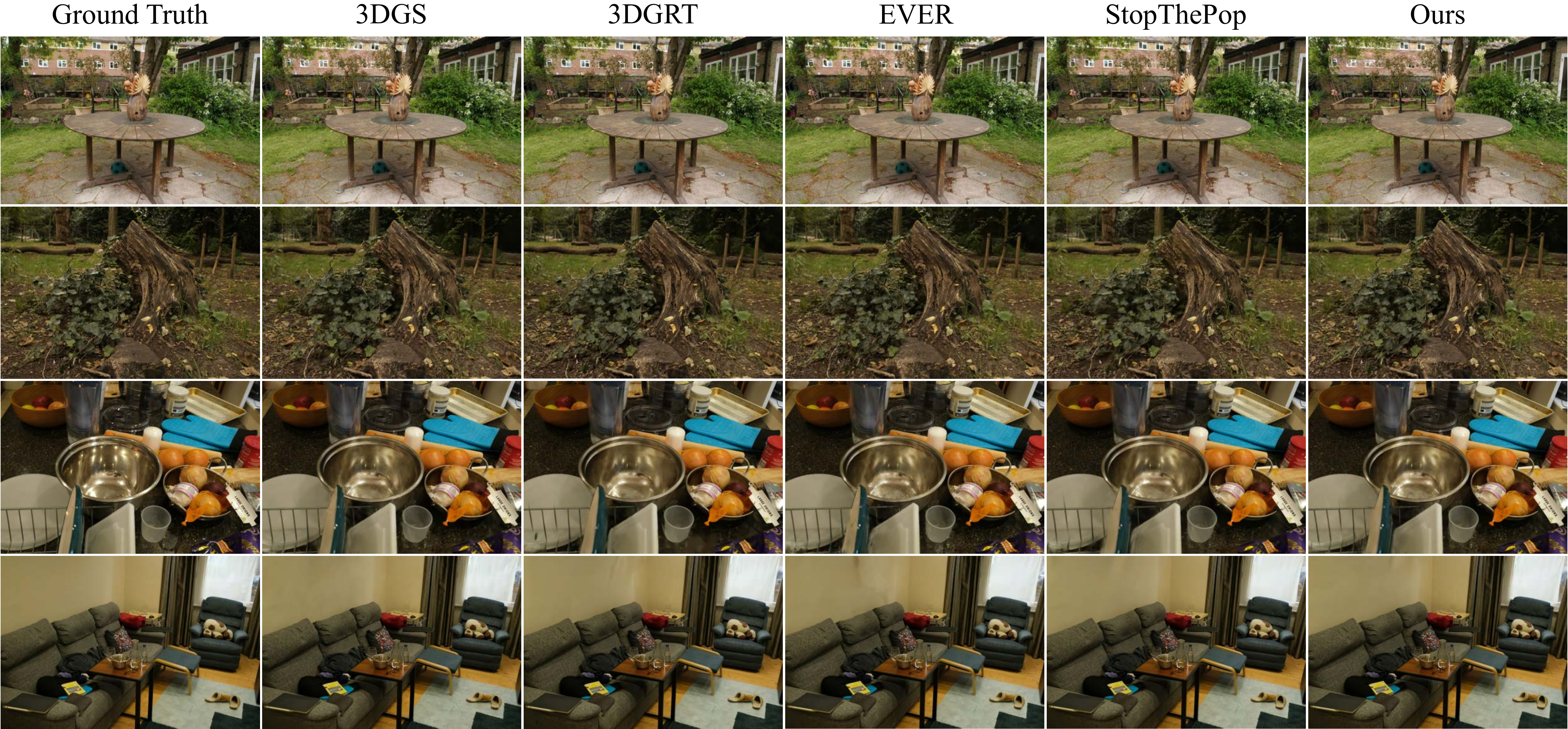}
    \vspace{-3mm}
    \caption{Qualitative comparison of our novel-view synthesis results against the baselines on the MipNERF360 dataset~\cite{barron2022mipnerf360}.}
    \label{fig:results_gallery}
\end{figure*}

\begin{table*}
\vspace{-2mm}
\caption{Quantitative results of our approach and baselines on the MipNERF360~\cite{barron2022mipnerf360} and Tanks 
\& Temples~\cite{Knapitsch2017} datasets.}
\vspace{-1mm}
\label{tab:main_benchmark}
\footnotesize
\vspace{-5mm}
\begin{center}
\begin{tabular}{l|cc|cccc|cccc}
\toprule
\multirow{2}{*}{Method\textbackslash{}Metric} & \multirow{2}{*}{\parbox{1cm}{Complex \\ Cameras}} & \multirow{2}{*}{\parbox{1cm}{Without \\ Popping}} & \multicolumn{4}{c|}{\texttt{MipNeRF360}} & \multicolumn{4}{c}{\texttt{Tanks \& Temples}} \\
 &  &  & \multicolumn{1}{c}{PSNR$\uparrow$} & \multicolumn{1}{c}{SSIM$\uparrow$} & \multicolumn{1}{c}{LPIPS$\downarrow$} & \multicolumn{1}{c|}{FPS $\uparrow$} & \multicolumn{1}{c}{PSNR$\uparrow$} & \multicolumn{1}{c}{SSIM$\uparrow$} & \multicolumn{1}{c}{LPIPS$\downarrow$} & \multicolumn{1}{c}{FPS $\uparrow$} \\ 
\midrule
ZipNeRF~\cite{barron2023zipnerf} & \checkmark & / & 28.54 & 0.828 & 0.219 & 0.2 & / & / & / & /\\
\midrule
3DGS~\cite{kerbl3Dgaussians} & \xmark  & \xmark  & \cellcolor{table_orange} 27.26 & 0.803 & 0.240 & \cellcolor{table_red}  347 & \cellcolor{table_red}  23.64 & \cellcolor{table_yellow}0.837 &  0.196 & \cellcolor{table_orange}  476 \\
Ours &  \checkmark  & \xmark  & \cellcolor{table_orange} 27.26 & 0.810 & \cellcolor{table_orange}0.218 & \cellcolor{table_yellow}265 & \cellcolor{table_orange}23.21 &\cellcolor{table_orange} 0.841 & \cellcolor{table_orange}0.178 & \cellcolor{table_yellow}277 \\
\midrule
StopThePop ~\cite{radl2024stopthepop} &  \xmark  & \checkmark & 27.14 & 0.804 & 0.235 & \cellcolor{table_orange}340 & 23.15 & \cellcolor{table_yellow}0.837 &  \cellcolor{table_yellow}0.189 & \cellcolor{table_red}482 \\
3DGRT~\cite{3dgrt2024} &  \checkmark  & \checkmark & \cellcolor{table_yellow} 27.20 & \cellcolor{table_orange} 0.818 & 0.248  & 52 & \cellcolor{table_yellow}23.20 & 0.830 & 0.222 & 190  \\
EVER~\cite{mai2024ever} &  \checkmark  & \checkmark &  \cellcolor{table_red} 27.51 & \cellcolor{table_red}  0.825 & \cellcolor{table_yellow}0.233 & 36 & / & / & / & / \\
Ours (sorted) &  \checkmark  & \checkmark & \cellcolor{table_orange} 27.26 & \cellcolor{table_yellow} 0.812 & \cellcolor{table_red}  0.215 & 200 & 22.90 & \cellcolor{table_red}  0.844 & \cellcolor{table_red}  0.172 & 272  \\
\bottomrule
\end{tabular}
\end{center}
\vspace{-6mm}
\end{table*}

\begin{table}
\caption{Detailed timings on the MipNeRF360~\cite{barron2022mipnerf360} dataset}
\label{tab:timings}
\footnotesize
\vspace{-7mm}
\begin{center}
\begin{tabular}{l|ccccc}
\toprule
Timings in ms & Preprocess & Duplicate & Sort & Render & Total \\
\midrule
3DGS~\cite{kerbl3Dgaussians}     &  0.59 & 0.34  & 0.55  & 1.27  & 2.88 \\
Ours & 1.34  & 0.31  & 0.33  & 1.61  & 3.77 \\
\midrule
StopThePop~\cite{radl2024stopthepop} & 0.57  & 0.27  & 0.14  & 1.83  & 2.94 \\
3DGRT~\cite{3dgrt2024} & /  & /  & /  & 19.24  & 19.24 \\
Ours (sorted) & 1.24  & 0.47  & 0.24  & 2.85  & 4.98   \\
\bottomrule
\end{tabular}
\end{center}
\vspace{-8mm}
\end{table}

\section{Experiments and Ablations}

In this section, we first evaluate the proposed approach on standard novel-view synthesis benchmark datasets~\cite{barron2022mipnerf360, Knapitsch2017}, analyzing both quality and speed. We additionally evaluate our method on an indoor dataset captured with fisheye cameras~\cite{yeshwanthliu2023scannetpp}, as well as an autonomous driving dataset captured using distorted cameras with rolling shutter effect~\cite{Sun_2020_waymo}. Ablation studies on key design choices and additional details on experiments and implementation are provided in the Supplementary Material.

\parahead{Model Variants} In the following evaluation, we will refer to two variants of our method. We use \emph{Ours} to denote the version that extends 3DGS~\cite{kerbl3Dgaussians} with the UT formulation (\cref{sec:unscented_tranform}) and particle evaluation in 3D (\cref{sec:evaluating_particle_response}). The second variant \emph{Ours (sorted)} additionally uses the per-ray sorting strategy as detailed in \cref{sec:unification} that leads to unification with 3DGRT~\cite{3dgrt2024} .

\parahead{Metrics}
We evaluate the perceptual quality of the novel views using peak signal-to-noise ratio (PSNR), learned perceptual image patch similarity (LPIPS), and structural similarity (SSIM) metrics.  To assess performance, we measure the time required for rendering a single image, excluding any overhead from data storage or visualization. For all evaluations, we use the datasets' default resolutions and report frames per second (FPS) measured on a single NVIDIA RTX 6000 Ada GPU.

\parahead{Baselines}
There have been many follow up works that improve or extend 3DGS in different aspects~\cite{kheradmand20243DGSMCMC, hahlbohm2024efficientperspectivecorrect3dgaussian, Yu2024MipSplatting, Condor2024Gaussians, scaffoldgs}. Many of these improvements are compatible with our approach, so we limit our comparison to the original 3DGS~\cite{kerbl3Dgaussians} and StopThePop~\cite{radl2024stopthepop} as the representative splatting methods, along with 3DGRT~\cite{3dgrt2024} and EVER~\cite{mai2024ever} as volumetric particle tracing methods that natively support distorted cameras and secondary lighting effects. On the dataset captured with fisheye cameras, we compare our method to FisheyeGS~\cite{liao2024fisheye} which extended 3DGS to fisheye cameras by deriving the Jacobian of the equidistant fisheye camera model. In addition to volumetric particle-based methods, we also compare our approach to state-of-the-art NeRF method ZipNeRF~\cite{barron2023zipnerf}.

\subsection{Novel View Synthesis Benchmarks}

\parahead{MipNeRF360~\cite{barron2022mipnerf360}} is the most popular novel-view synthesis benchmark consisting of nine large scale outdoor and indoor scenes. Following prior work, we used the images downsampled by a factor of four for the outdoor scenes, and by a factor of two for the indoor scenes. To enable comparison with other splatting method, we use rectified images provided by \citet{kerbl3Dgaussians}.

\cref{tab:main_benchmark} depicts the quantitative comparison, while the qualitative comparison on selected scenes is provided in Fig. \ref{fig:results_gallery}. As anticipated, on this dataset with perfect pinhole inputs, both \emph{Ours} and \emph{Ours (sorted)} achieve comparable perceptual quality to other splatting and tracing methods.  
In terms of inference runtime~\cref{tab:main_benchmark}, our method achieves comparable frame rates to 3DGS~\cite{kerbl3Dgaussians}, while greatly outperforming all other methods that support complex cameras at more than 265FPS while the closest competitor, 3DGRT~\cite{3dgrt2024}, achieves 52FPS. 

\parahead{Tanks \& Temples~\cite{Knapitsch2017}}  contains two large-scale outdoor scenes where the camera circulates around a prominent object (\emph{Truck} and \emph{Train}). Both scenes include lighting variations, and the \emph{Truck} scene also contains transient objects that should ideally be ignored by reconstruction methods. \cref{tab:main_benchmark} depicts the quantitative comparison while the qualitative results are provided in the Supplementary Material.

\parahead{Scannet++~\cite{yeshwanthliu2023scannetpp}}  is a large-scale indoor dataset captured with a fisheye camera at a resolution of 1752 × 1168 pixels. For our evaluation, we use the same six scenes as FisheyeGS~\cite{liao2024fisheye} and follow the same pre-processing steps. Specifically, we convert the images to an equidistant fisheye camera model to match the requirements of~\cite{liao2024fisheye}. \footnote{Note that our method seamlessly supports the full fisheye camera model without any code modifications.}

\begin{figure}
    \centering
    \includegraphics[width=\linewidth]{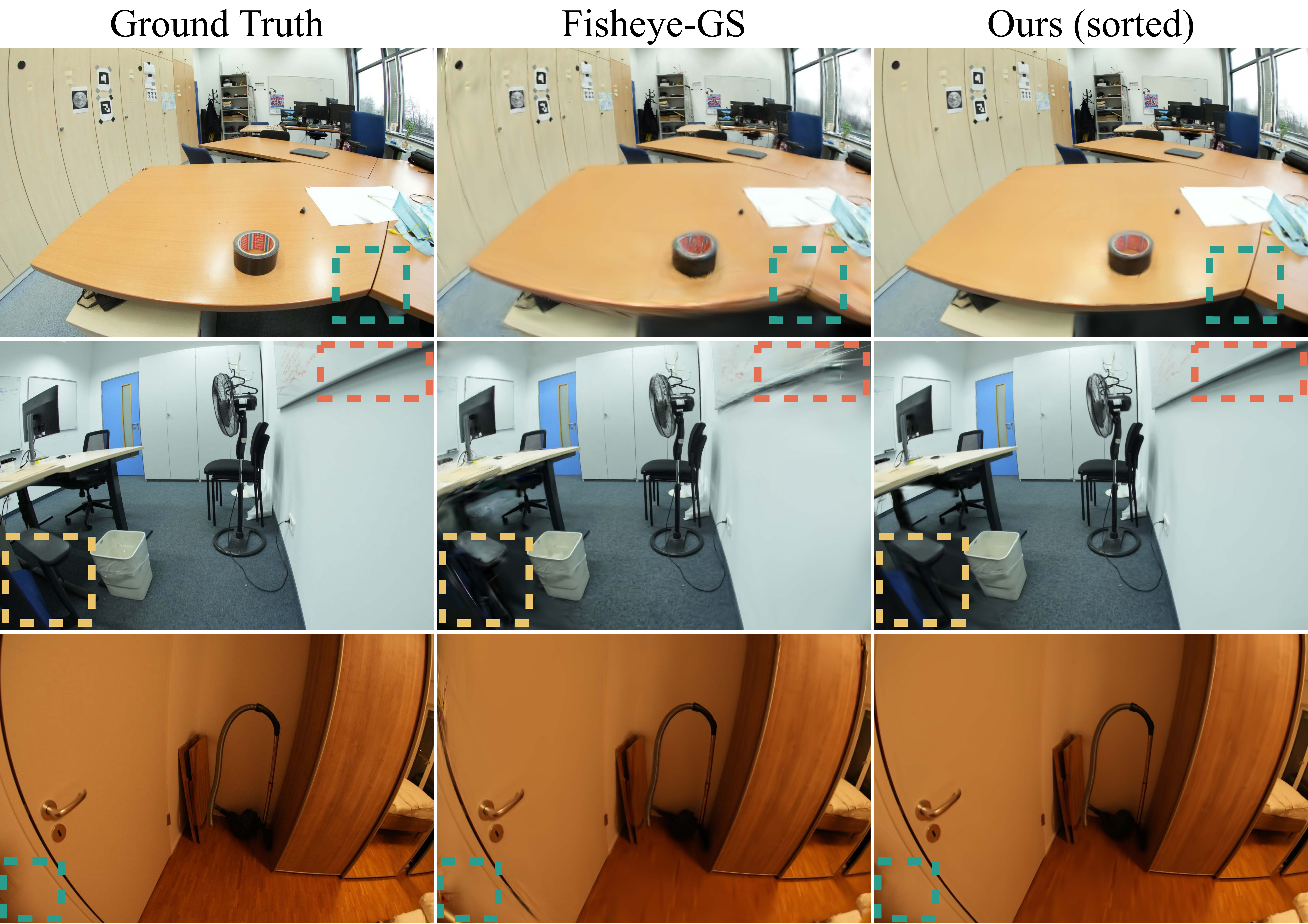}
    \vspace{-7mm}
    \caption{Comparison of our renderings against Fisheye-GS~\cite{liao2024fisheye}, on scenes from the Scannet++ dataset~\cite{yeshwanthliu2023scannetpp}.}
    \label{fig:scannet_comparison}
    \vspace{-2mm}
\end{figure}

On this dataset, we compare \emph{Ours} to FisheyeGS~\cite{liao2024fisheye} and 3DGS~\cite{liao2024fisheye}. The results for the latter are taken from~\cite{liao2024fisheye} where they were obtained by: (i) undistorting the training images and training with the official 3DGS~\cite{kerbl3Dgaussians} implementation, and (ii) rendering  equidistant fisheye test views from that representation using the FisheyeGS~\cite{liao2024fisheye} formulation. This setting is unfavorable for 3DGS~\cite{liao2024fisheye} as significant portions of the images are lost during undistortion, but it highlight the problem of being limited to perfect pinhole cameras. The quantitative comparison is shown in ~\cref{tab:fisheye_gs_benchmark} and qualitative results are provided in ~\cref{fig:scannet_comparison}. \emph{Ours} significantly outperforms FisheyeGS~\cite{liao2024fisheye} across all perceptual metrics, while using less than half the particles (1.07M vs. 0.38M). This result underscores the flexibility and potential of our approach. Despite FisheyeGS~\cite{liao2024fisheye} deriving a Jacobian for this particular camera model---limiting its applicability even to similar models (e.g., fisheye with distortions)---it still underperforms our simple formulation that can be trivially applied to any camera model. 

\begin{table}

\caption{When evaluated on a dataset acquired with equidistant fisheye cameras, our general method outperforms~\cite{liao2024fisheye} which derived the linerization for this specific camera model. Undistortion removes large parts of the original images and results in underobserved regions~\cite{kerbl3Dgaussians}. Results marked with $\dagger$ are taken from~\cite{liao2024fisheye}.}
\vspace{-6mm}
\label{tab:fisheye_gs_benchmark}
\footnotesize
\begin{center}
\resizebox{\columnwidth}{!}{
\begin{tabular}{l|cccc}
\toprule
\multirow{2}{*}{Method\textbackslash{}Metric} & \multicolumn{4}{c}{\texttt{Scannet++}} \\
 & \multicolumn{1}{c}{PSNR$\uparrow$} & \multicolumn{1}{c}{SSIM$\uparrow$} & \multicolumn{1}{c}{LPIPS$\downarrow$} & \multicolumn{1}{c}{N. Gaussians$\downarrow$} \\
\midrule
3DGS$^{\dagger}$ & 22.76 & 0.798 & / & 1.31M \\
FisheyeGS$^{\dagger}$~\cite{liao2024fisheye} & \cellcolor{table_yellow} 27.86 & \cellcolor{table_yellow} 0.897 & / & \cellcolor{table_yellow} 1.25M \\
FisheyeGS~\cite{liao2024fisheye} & \cellcolor{table_orange} 28.15 & \cellcolor{table_orange} 0.901 & \cellcolor{table_orange} 0.261 & \cellcolor{table_orange} 1.07M \\
Ours (sorted) & \cellcolor{table_red} 29.11 & \cellcolor{table_red}0.910 & \cellcolor{table_red} 0.252 & \cellcolor{table_red} 0.38M \\
\bottomrule
\end{tabular}
}
\end{center}
\vspace{-6mm}
\end{table}

\begin{figure}
    \centering
    \includegraphics[width=\linewidth]{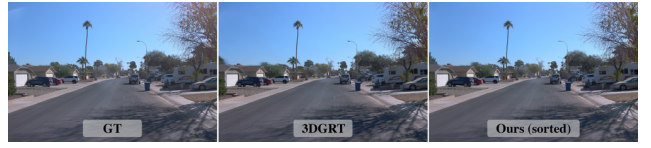}
    \vspace{-7mm}
    \caption{Qualitative comparison of our novel-view synthesis results against 3DGRT on the Waymo dataset~\cite{Sun_2020_waymo}.}
    \label{fig:waymo}
    \vspace{-2mm}
\end{figure}

\parahead{Waymo~\cite{Sun_2020_waymo}} is a large scale autonomous driving dataset captured using distorted cameras with rolling-shutter. We follow 3DGRT~\cite{3dgrt2024} and select 9 scenes with no dynamic objects to ensure accurate reconstructions. \cref{fig:waymo} show qualitative results. \emph{Ours (sorted)} can faithfully represent complex camera mounted on a moving platform and reaches comparable performance to 3DGRT~\cite{3dgrt2024}. More \revRemoved{quantitative and qualitative} results are provided in the Supplementary Material.

\begin{figure}
    \def\svgwidth{1.\linewidth}
    \centering
    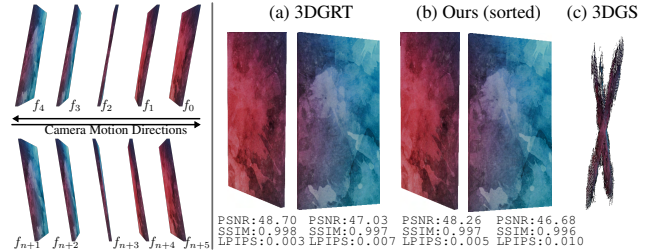
    \vspace{-6mm}
    \caption{
        Multiple frame tiles $f_i$ of a
        \emph{single} solid box rendered by a left- and right-panning rolling
        shutter camera with a top-to-bottom shutter direction illustrate this
        \emph{time-dependent} sensor effect (data from \cite{3dgrt2024}).
        While ray-tracing-based methods like 3DGRT naturally support
        compensating for these time-dependent effects (a), traditional splatting
        methods struggle to model these (c), whereas our UT-based
        splatting formulation faithfully incorporates the sensor's motion into the
        projection formulation and recuperates the true undistorted geometry
        (b).
        }
    \label{fig:RS-cameras}
\end{figure}

\section{Applications}
\ourmethod also enables novel applications and techniques that were previously unattainable with particle scene representation within a rasterization framework.

\subsection{Complex cameras}

\begin{figure}
    \centering
    \includegraphics[width=\linewidth]{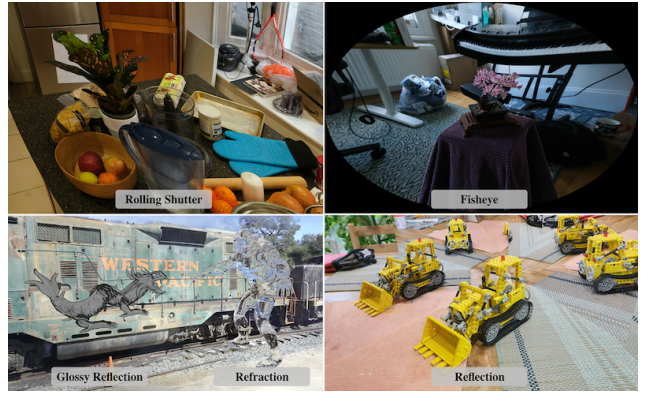}
    \vspace{-7mm}
    \caption{Illustration of the effects unlocked by our method. Top-left: rendering an image with rolling-shutter. Top-right : applying a strong lens distortion. Bottom : hybrid splatting / tracing rendering. Primary rays are splatted using our method while secondary rays are traced using 3DGRT \cite{3dgrt2024}. This hybrid formulation allows us to simulate refraction (left) and reflections (right).}
    \label{fig:rendering_playground}
    \vspace{-4mm}
\end{figure}

\parahead{Distorted Camera Models} Projection of particles using UT enables \ourmethod not only to train with distorted cameras, but also to render different camera models with varying distortion from scenes that were trained using perfect pinhole camera inputs (\cref{fig:rendering_playground} top row).

\parahead{Rolling Shutter} Apart from the modeling of distorted cameras, \ourmethod can also faithfully incorporate the camera motion into the projection formulation, hence offering support for time-dependent camera effects such as rolling-shutter, which are commonly encountered in the fields of autonomous driving and robotics. Although optical distortion can be addressed with image rectification\footnote{Image rectification is generally effective only for low-FoV cameras and results in information loss, as shown in \cref{tab:fisheye_gs_benchmark}.}, incorporating time-dependency of the projection function in the linearization framework is highly non-trivial. 

To illustrate the impact of rolling shutter on various reconstruction methods, in \cref{fig:RS-cameras} we use the synthetic dataset provided by \citet{3dgrt2024} where the motion of the camera and the shutter time are provided.

\begin{figure}
    \centering
    \includegraphics[width=\linewidth]{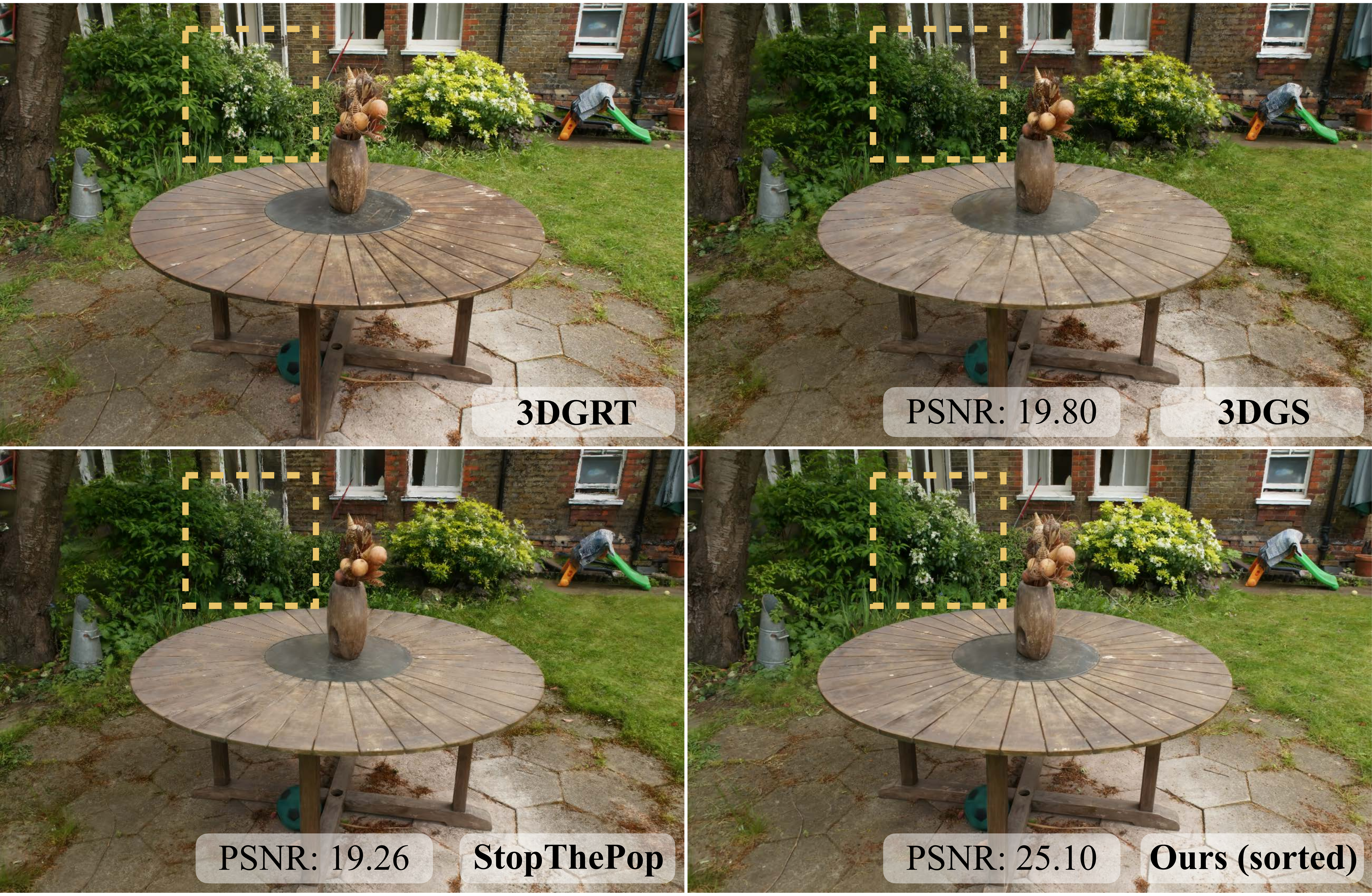}
    \vspace{-7.5mm}
    \caption{Scenes trained with different methods and rendered using 3DGRT~\cite{3dgrt2024}. Our method is the most consistent with the tracing approach, allowing for seamless hybrid rendering with splatting for primary and tracing for secondary rays.}
    \label{fig:rendering_ablation}
    \vspace{-2mm}
\end{figure}

\subsection{Secondary rays and lighting effects}

\parahead{Aligning the representation with 3DGRT~\cite{3dgrt2024}} The rendering formulations of 3DGS and 3DGRT mainly differ in terms of (i) determining which particles contribute to which pixels, (ii) the order of particles evaluation, \revAdded{and} (iii) the computation of the particles response. In \cref{sec:evaluating_particle_response,sec:unification}\revAdded{,} our goal was to reduce these differences to arrive \revReplaced{to}{at} a common 3D representation that can be both rasterized and traced. ~\cref{fig:rendering_ablation} shows the comparison of 3D representations trained with different methods and evaluated with 3DGRT~\cite{3dgrt2024}. While some discrepancies naturally remain, \emph{Ours (sorted)} achieves much better alignment \revReplaced{to}{with} 3DGRT than StopThePop or 3DGS.

\parahead{Secondary rays} Aligning our rendering formulation to 3DGRT~\cite{3dgrt2024} enables \revRemoved{us to perform }hybrid rendering by rasterizing the primary and tracing the secondary rays within the same representations. Specifically, we first compute all the primary rays intersections with the scene, then \revReplaced{we can have these primary rays rendered}{render these primary rays} using \revReplaced{our splatting method by simply discarding}{rasterization and discard} Gaussian hits \revReplaced{falling}{that fall} behind a ray's closest intersection. Next, we compute \revReplaced{secondary rays and trace them}{and trace the secondary rays} using 3DGRT. This hybrid rendering method allows us to achieve \revRemoved{most of the }complex visual effects, such as reflections and refractions, that would otherwise only be possible with ray tracing.

\vspace{-2mm}
\section{Discussion}
We proposed a simple idea to replace the linearization of the non-linear projection function in 3DGS~\cite{kerbl3Dgaussians} with the Unscented Transform. This modification enables us to seamlessly generalize 3DGS to distorted cameras, support time-dependent effects such as rolling shutter, and align our rendering formulation with 3DGRT~\cite{3dgrt2024}. The latter enables us to perform hybrid rendering and unlock secondary rays for lighting effects.

\parahead{Limitations and Future Work} Our method is significantly more efficient than ray-tracing-based methods~\cite{3dgrt2024, Condor2024Gaussians, mai2024ever}, but it is still marginally slower than~\cite{kerbl3Dgaussians} (see details in ~\cref{tab:timings}). While being more general, the UT evaluation and the added complexity of 3D particle evaluation impact rendering times. 
Additionally, although UT permits exact projection of sigma points under arbitrary distortions, the resulting projected shape deviates from a 2D Gaussian in case of large distortions. This degrades the approximation of which particles contribute to which pixels. Finally, as our method still uses a single point to evaluate each primitive, it is currently unable to render overlapping Gaussians accurately. Approaches such as EVER~\cite{mai2024ever} may offer promising directions for addressing this limitation. Looking ahead, we hope that this work could inspire new research, particularly in fields like autonomous driving and robotics, where training and rendering with distorted cameras is essential. Our alignment with 3DGRT~\cite{3dgrt2024} also opens interesting opportunities for future research in inverse rendering and relighting.

\section{Acknowledgements}

We thank our colleagues Riccardo De Lutio, Or Perel, and Nicholas Sharp for their help in setting up experiments and for their valuable insights that helped us improve this work.

\clearpage
\setcounter{section}{0} %
\renewcommand{\thesection}{\Alph{section}} %
\setcounter{page}{1}
\maketitlesupplementary

\noindent
In this supplementary material, we present an extension to generalized Gaussian
particles (\cref{sec:generalized_gaussian_particles}), derive a numerically stable scheme for computing the partial derivative through the proposed 3D particle evaluation (\cref{sec:derivative}, cf. \cref{sec:evaluating_particle_response}), and provide
further ablations of the proposed UT-based rasterization (\cref{sec:projection_validation}). We also include details on autonomous vehicle dataset reconstructions (\cref{sec:waymo_eval}). Finally, we summarize the Gaussian rasterization algorithm  and demonstrate that our method serves as a drop-in replacement for a small part of it  (\cref{sec:algorithm}).

\section{Generalized Gaussian Particles}
\label{sec:generalized_gaussian_particles}
In 3DGRT~\cite{3dgrt2024} the authors propose to use particles with different kernel functions and their most efficient approach is based on a \emph{generalized Gaussians of degree $2$}. In~\cref{tab:generalized_particles} we demonstrate that our approach supports different particles as well. Different to~\cite{3dgrt2024}, we define a generalized Gaussians kernel function of degree $n$ as
\begin{equation}
     \ParticleResponse(\bm{x}) = \exp(-\lambda((\bm{x}-\ParticleCenter)^T\bm{\Sigma^{-1}}(\bm{x}-\ParticleCenter))^{\frac{n}{2}}) 
\end{equation}
with $\lambda=\frac{r^2}{r^n}$ a scale factor defined to get the same kernel response at a given distance $r$ as the reference Gaussian kernel (we use $r=3$). 
Note that 3DGRT \emph{generalized Gaussians of degree $2$} corresponds to our generalized Gaussians kernel of degree 4.
\begin{figure}[h!]
    \centering
    \includegraphics[width=\linewidth]{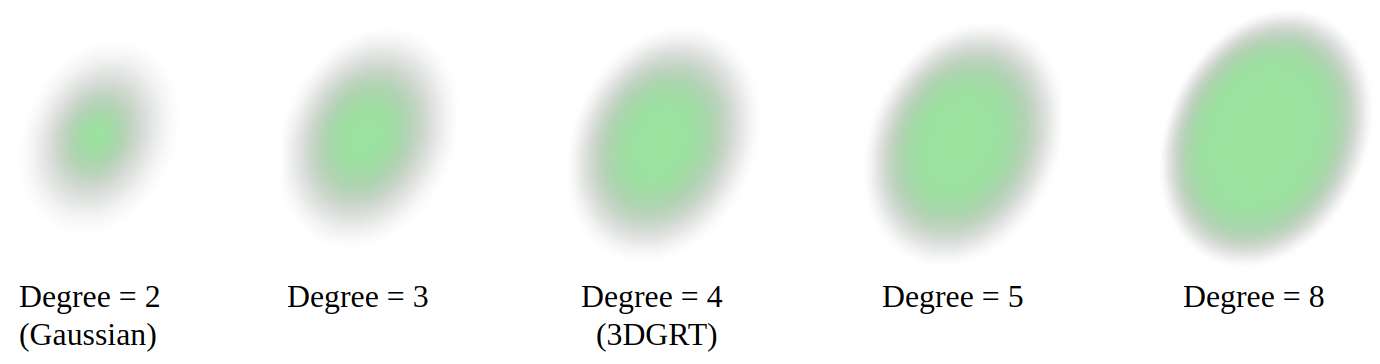}
    \caption{Rendering the same generalized Gaussian particle with different degrees. Higher degree particles are denser and have a steeper and narrower fall-off.}
    \label{fig:particle_generalized}
\end{figure}

\cref{fig:particle_generalized} illustrates the effect of using a different kernel function on the particle extent and density.
\begin{table}[]
\caption{
Quality and speed tradeoffs computed on MipNERF360~\cite{barron2022mipnerf360} (excluding \emph{flower} and \emph{treehill} for fair comparison with 3DGRT) for various particle generalized Gaussian kernel functions. Note that our kernel of degree$=4$ corresponds to the generalized Gaussian of degree$=2$ proposed in 3DGRT~\cite{3dgrt2024}.}
\footnotesize
\begin{center}
\begin{tabular}{l|ll|ll}
\toprule
 & \multicolumn{4}{c}{\texttt{MipNERF360}} \\
 & \multicolumn{2}{c|}{Ours (sorted)} & \multicolumn{2}{c}{3DGRT} \\
 Kernel function & \multicolumn{1}{c}{PSNR$\uparrow$} & \multicolumn{1}{c|}{FPS$\uparrow$} & \multicolumn{1}{c}{PSNR$\uparrow$} & \multicolumn{1}{c}{FPS$\uparrow$} \\ \midrule
Degree $=2$ (Gaussian) & 28.77  & 207 & 28.69 & 55 \\
Degree $=3$ & 28.71 & 217 & / & / \\
Degree $=4$ (3DGRT) & 28.46 & 233 & 28.71 & 78 \\
Degree $=5$ & 28.33 & 238 & / & / \\
Degree $=8$ &  27.63 & 243 & / & / \\
\bottomrule
\end{tabular}
\end{center}
\label{tab:generalized_particles}
\end{table}

\cref{tab:generalized_particles} shows how the degree of the generalized Gaussian kernel function permits to better control the trade-off between the rendering quality and speed.

\section{Derivation of Backward Gradients}
\label{sec:derivative}
In the following, we provide a step-by-step derivation of $\partial \alpha / \partial \ParticleCenter$. The derivations of $\partial \alpha / \partial \bm{S}$ and $\partial \alpha / \partial \bm{R}$ follow analogously.

\begin{figure}[t]
    \centering
    \includegraphics[width=\linewidth]{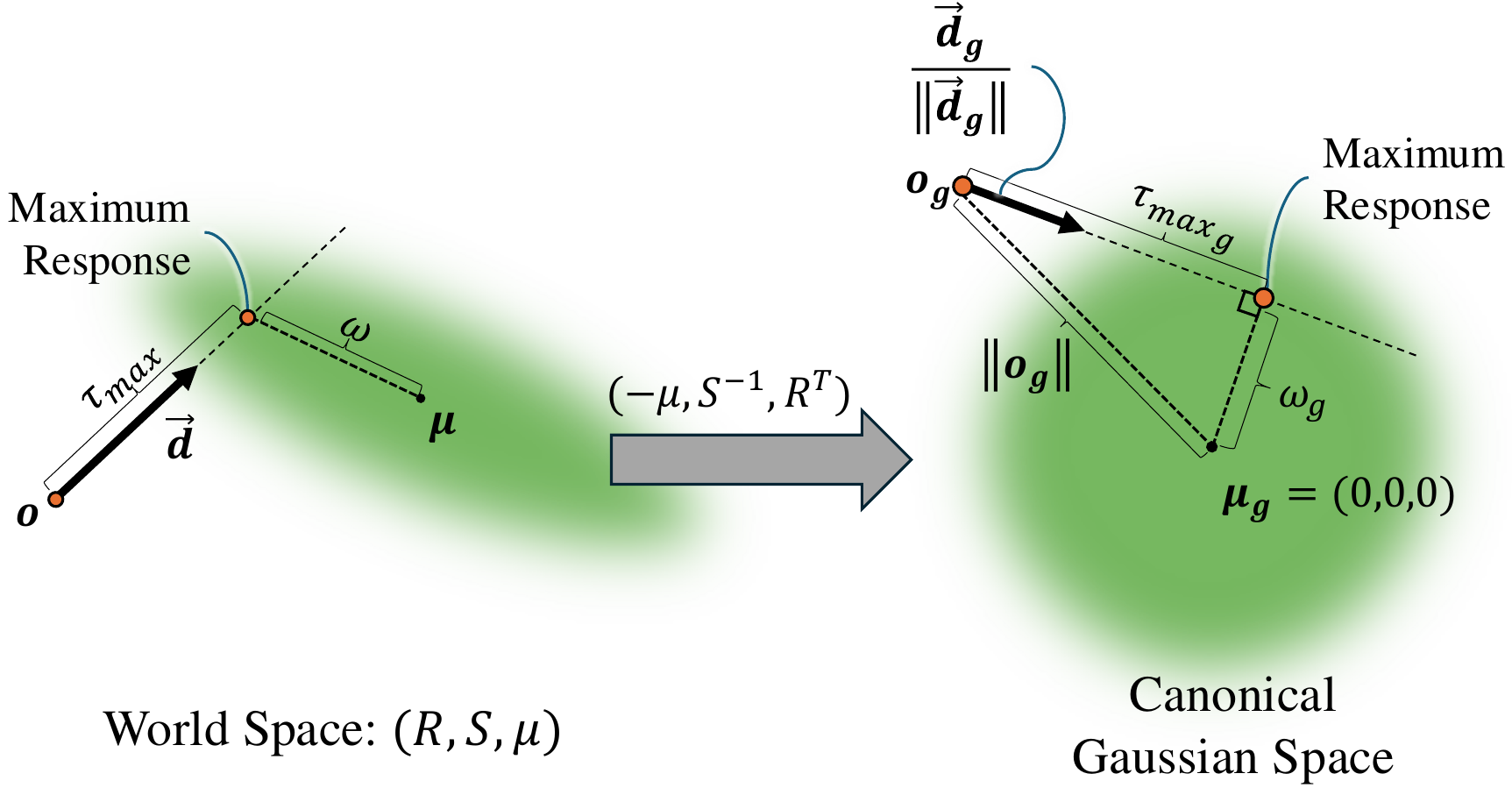}
    \vspace{-5mm}
    \caption{An illustration of the geometric transformation of a Gaussian from world space to canonical Gaussian space.}
    \label{fig:particle_response_grad}
\end{figure}

Remember that $\alpha =\sigma\ParticleResponse(\RayOrigin + \tau_{\text{max}}\RayDirection)$ and consider that $\tau_{\text{max}}$ can be defined in the canonical Gaussian space as
\begin{equation}
    \tau_{\text{max}_{g}} =-\RayOrigin_g^T\frac{\RayDirection_g}{||\RayDirection_g||},
\end{equation}
where $\RayOrigin_g=\bm{S}^{-1}\bm{R}^T(\RayOrigin-\ParticleCenter)$ and $\RayDirection_g=\bm{S}^{-1}\bm{R}^T\RayDirection$ denote the ray origin and ray direction expressed in Gaussian canonical space, respectively. An illustration of the geometric relationship between values is provided in \cref{fig:particle_response_grad}.

Let $\omega_g^2=||\RayOrigin_g + \tau_{\text{max}_g}\frac{\RayDirection_g}{||\RayDirection_g||}||^2$ denote the squared distance from the Gaussian particle center to the point of maximum response such that $\alpha = \sigma e^{-0.5\omega_g^2}$. The partial derivatives can be computed as

\begin{align}
    \dfrac{\partial \alpha}{\partial \omega_g^2} &= -0.5\sigma e^{-0.5\omega_g^2}\\
    \dfrac{\partial \omega_g^2}{\partial \RayOrigin_g} &= 2\RayOrigin_g + 2\tau_{max_g}\frac{\RayDirection_g}{||\RayDirection_g||}\\
    \dfrac{\partial \RayOrigin_g}{\partial \ParticleCenter} &= -\bm{S}^{-1}\bm{R}^T
\end{align}

\section{Gaussian Projection Quality}
\label{sec:projection_validation}

\begin{figure}
  \centering
  \includegraphics[width=\linewidth]{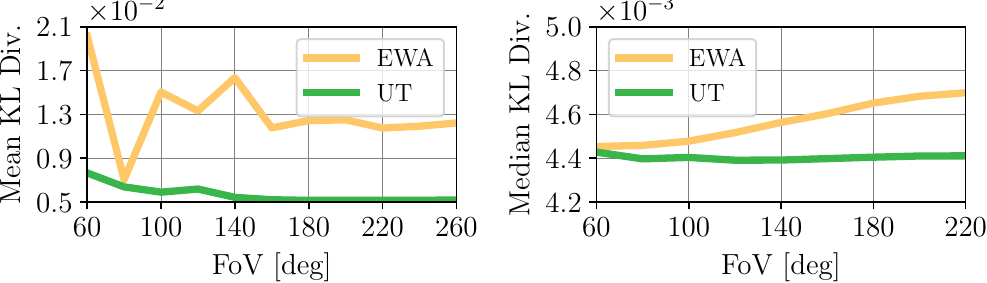}
  \vspace{-7mm}
  \caption{KL divergence to Monte Carlo for equidistant fisheye cameras.}
  \label{fig:equidistant_fisheye}
  \vspace{-2mm}
\end{figure}

\begin{figure}
  \centering
  \includegraphics[width=\linewidth]{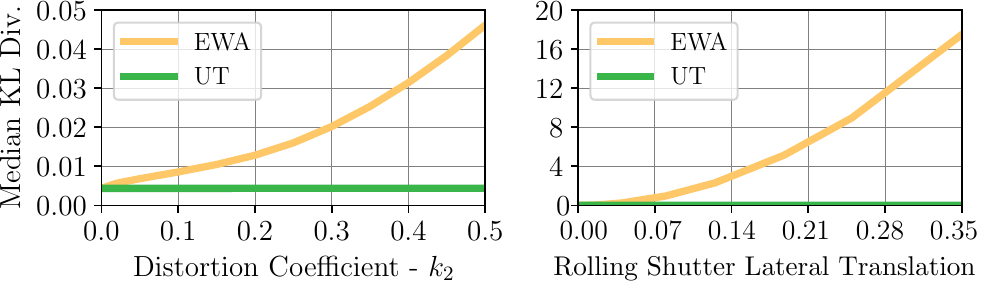}
  \vspace{-7mm}
  \caption{KL divergence to Monte Carlo under radial distortion and rolling shutter.}
  \label{fig:radial_distortion_rolling_shutter}
  \vspace{-2mm}
\end{figure}

\begin{figure*}
  \def\svgwidth{1.\linewidth}
  \centering
  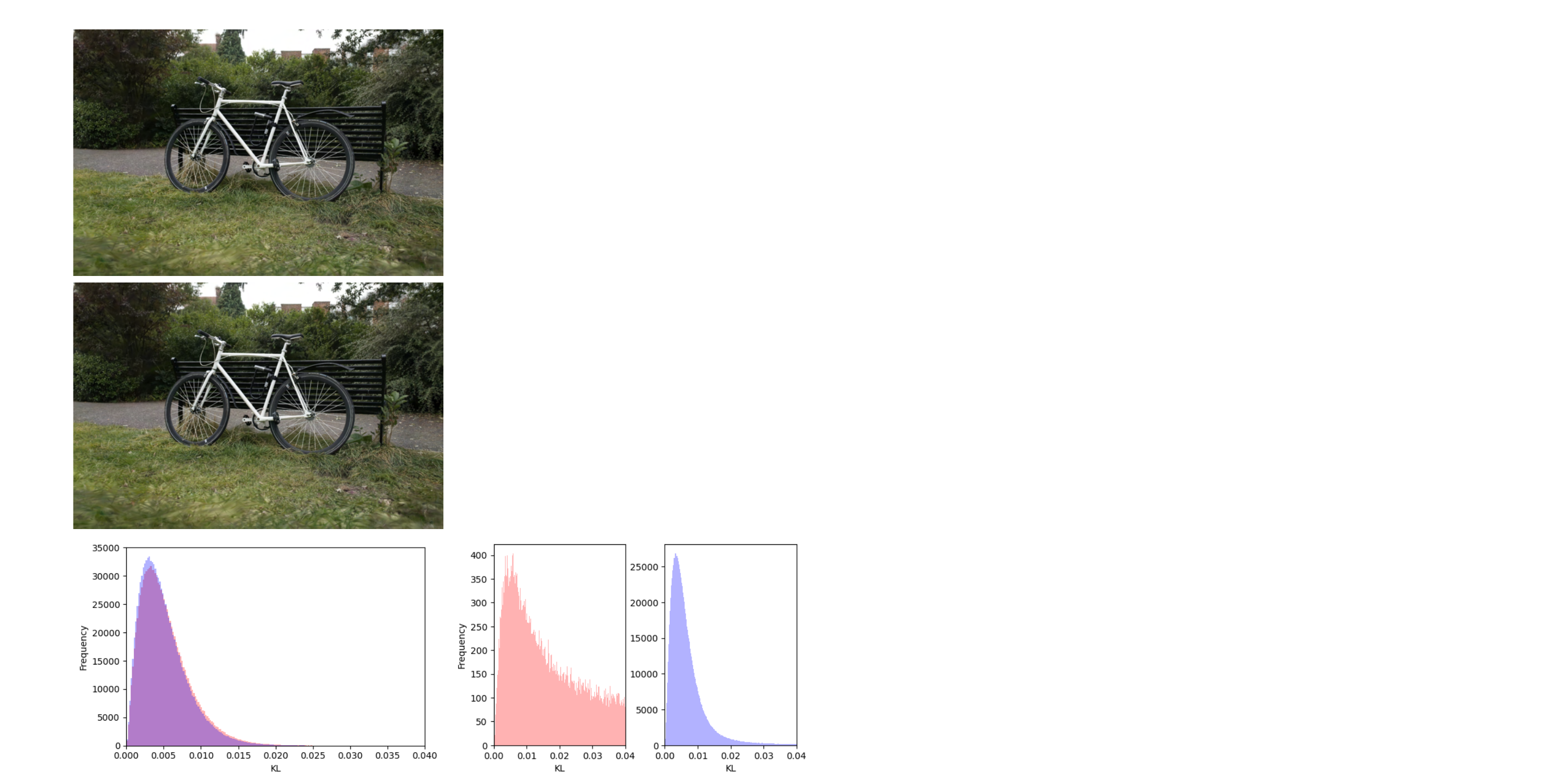
  \vspace{-6mm}
  \caption{
      Gaussian Projection Quality: for both distortion-free pinhole and fisheye
      camera models, as well as static and rolling-shutter (RS, top-top-bottom
      shutter direction) poses, we evaluate the Kullback–Leibler (KL
      $\downarrow$) divergence of each Gaussian projected using either EWA
      ({\color[RGB]{255,178,178} $\bullet$}) or UT-based
      ({\color[RGB]{178,178,255} $\bullet$}) projections against
      \emph{Monte-Carlo}-based reference projection.
      The distribution of KL-divergences for each rendering is shown in the
      histograms below
      }
  \label{fig:kl-dists}
\end{figure*}

While Monte Carlo sampling (cf. \cref{fig:UT}) is expensive to compute, it
provides accurate reference distributions for assessing the quality of both EWA and
the proposed UT-based projection methods.
This assessment can be quantified using the Kullback–Leibler (KL) divergence between both
2d distributions, where lower KL values indicate the projected Gaussians better
approximate the reference projections.
In \cref{fig:kl-dists}, we evaluate the KL divergence for a fixed reconstruction ({\small\texttt{MipNERF360 bicycle}~\cite{barron2022mipnerf360}}). Specifically, for each visible Gaussian, we compare the projections obtained using either method under different camera and pose configurations against MC-based references (using 500 samples per reference).
The resulting KL divergence distributions are visualized in the histograms at the bottom.

While both distributions of divergences are consistent for the static pinhole
camera case (first column), UT-based projections are more accurate compared to
EWA-based estimates for the static fisheye camera case (third column),
indicating that UT yields a better approximation in case of higher non-linearity of the projection.
For rolling-shutter \revAdded{(RS)} camera poses (second and fourth columns), RS-aware UT-based
projections still approximate the RS-aware MC references well. In contrast,
RS-unaware EWA linearizations break down and fail to approximate this case
(histogram domains are capped to $0.04$ for clearer visualization, but the
EWA-based projections have a long tail distribution of larger KL values
still).
The tearing artifacts observed in EWA-based RS renderings arise from these inaccurate projections, leading to incorrect pixel-to-Gaussian associations during the volume rendering step.

\revAdded{Additionally, we provide quantitative evaluation of distortion effects.
\cref{fig:equidistant_fisheye} further illustrates the KL divergence relative to MC projection across different FoV using an equidistant fisheye camera model. Our approach provides more accurate approximations than even the custom-derived Jacobian employed for EWA splatting.
\cref{fig:radial_distortion_rolling_shutter} shows the same comparison under increasing radial distortion and RS. For EWA we use the Jacobian from~\cite{kerbl3Dgaussians}, which does not account for these additional distortions. While one could derive a custom Jacobian for radial distortion, linearizing the RS effect is non-trivial. In contrast, our general UT-based method maintains virtually the same median KL divergence regardless of the distortion parameter $k_2=0.0$ $(\text{KL}_{\text{median}}=4.4\times10^{-3})$ and $k_2=0.5$ $(\text{KL}_{\text{median}}=4.3\times10^{-3})$ and similarly remains consistent under RS lateral translations of $0.0$ $(\text{KL}_{\text{median}}=4.4\times10^{-3})$ and $0.35$ $(\text{KL}_{\text{median}}=4.6\times10^{-3})$.}

\begin{table}[t]
\caption{On the Waymo~\cite{Sun_2020_waymo} autonomous vehicles dataset that was captured with distorted camera model and rolling-shuter sensor, our method achieves better quality compared to 3DGRT~\cite{3dgrt2024}. Note that 3DGS~\cite{kerbl3Dgaussians} requires the training and evaluation to be done on rectified images without rolling shutter effects and is hence not directly comparable.}
\label{tab:waymo_benchmark}
\vspace{-5mm}
\footnotesize
\begin{center}
\begin{tabular}{l|cc}
\toprule
\multirow{2}{*}{Method\textbackslash{}Metric} & \multicolumn{2}{c}{\texttt{Waymo}} \\
 & \multicolumn{1}{c}{PSNR$\uparrow$} & \multicolumn{1}{c}{SSIM$\uparrow$} \\
\midrule
3DGS~\cite{kerbl3Dgaussians} &                          29.83 & \cellcolor{table_red}    0.917 \\
\midrule
3DGRT~\cite{3dgrt2024}       & \cellcolor{table_yellow} 29.99 &                          0.897 \\
Ours (sorted)                & \cellcolor{table_red}    30.16 & \cellcolor{table_yellow} 0.900 \\
\bottomrule
\end{tabular}
\end{center}
\vspace{-5mm}
\end{table}

\section{Waymo Autonomous Vehicle Dataset}
\label{sec:waymo_eval}
\begin{figure*}
    \centering
    \includegraphics[width=\linewidth]{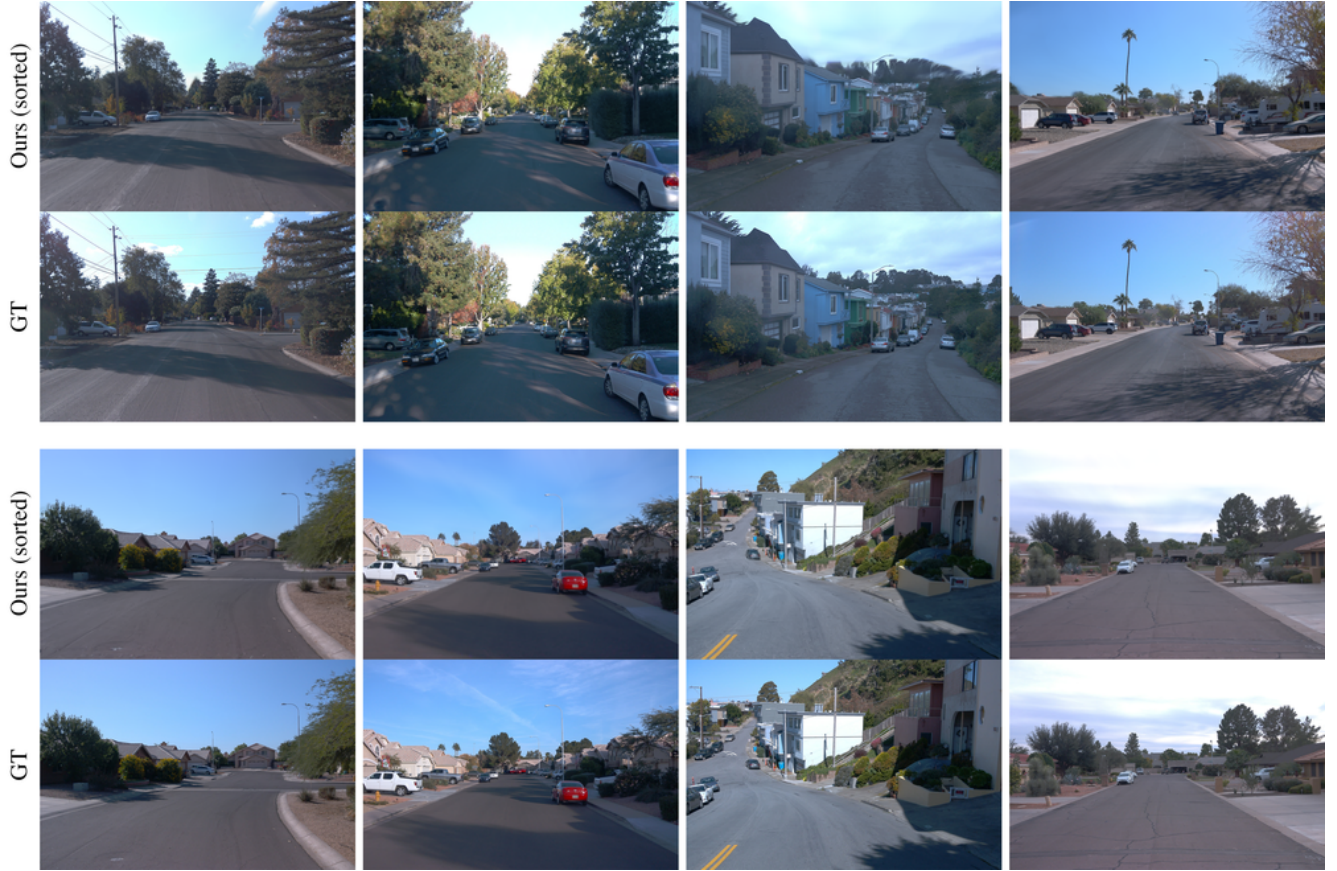}
    \caption{Qualitative comparison of our novel-view synthesis results against the ground truth on the Waymo dataset~\cite{Sun_2020_waymo}. Images are sampled from 8 different scenes.}
    \label{fig:waymo_image}
\end{figure*}

For comparison on the Waymo Open Perception dataset~\cite{Sun_2020_waymo}, we follow \cite{3dgrt2024} and select 9 static scenes. Images in the dataset are captured using a distorted camera with rolling shutter sensor, mounted on the front of the vehicle. 
To adapt to this dataset, we incorporated additional losses for lidar depth and image opacity, combining them as a weighted sum: the L1-loss $\mathcal{L}^{\text{depth}}_1$ for depth and the L2-loss $\mathcal{L}^{\text{opacity}}_2$ for opacity, such that $\mathcal{L}^{\text{waymo}} = \mathcal{L} + 0.001 \mathcal{L}^{\text{depth}}_1 + 0.05 \mathcal{L}^{\text{opacity}}_2$, where $\mathcal{L}$ is the loss function defined in \cref{sec:training}.
We initialized scenes using a colored point cloud generated by combining screen-projected lidar points with camera data.
For the case of 3DGS~\cite{kerbl3Dgaussians}, we rectify the images and ignore the rolling shutter effects following~\cite{chen2024omnire}. For 3DGRT~\cite{3dgrt2024} and our method, we make use of the full camera model and compute the rolling shutter effect correctly. The quantative results are reported in \cref{tab:waymo_benchmark} and qualitative visualizations are available in \cref{fig:waymo_image}.

\noindent
\begin{minipage}{\columnwidth} %
\vspace{2em}
  \begin{algo}{\Proc{Rasterize}}
  \label{proc:rasterize}
  \begin{algorithmic}[1]
    \InputConditions{Gaussian parameters: $\{\ParticleCenter_i, \bm{R}_i, \bm{S}_i, \sigma_i\}_{i=1}^N$,\\ 
                     camera extrinsic $\bm{W}$, camera intrinsic $\bm{D}$ }
    \OutputConditions{2D Means: $\bm{v}_{\ParticleCenter_i}$, 2D AABBs: $\bm{r}_i$}
    \For{$i \text{ in } 1 \dots N$} \Comment{ iterate over the particles}
    \State $\bm{v}_{\ParticleCenter_i}, \bm{\Sigma'}_i = \text{Estimate2DGaussian}(\ParticleCenter_i, \bm{R}_i, \bm{S}_i, \bm{W}, \bm{D})$
    \State $\bm{h}_i = \text{Extent}(\bm{\Sigma'}_i, \sigma_i)$ 
    \State \Comment{ use opacity to compute a tighter 2D extent}
    \State $\bm{r}_i = \text{ComputeRectangle}(\bm{h}_i,\bm{v}_{\ParticleCenter_i})$ 
    \State \Comment{ 2D rectangle used for tile-based rasterization}
    \EndFor
  \end{algorithmic}
\end{algo}
\end{minipage}

\noindent
\begin{minipage}{\columnwidth} %
\vspace{2em}
  \begin{algo}{\Proc{Estimate2DGaussian}}
  \label{proc:UT}
  \begin{algorithmic}[1]
    \InputConditions{Gaussian parameters: $\ParticleCenter, \bm{R}, \bm{S}$,\\
                     camera extrinsic $\bm{W}$, camera intrinsic $\bm{D}$, $\alpha$, $\beta$, $\kappa$}
    \OutputConditions{2D Mean: $\bm{v}_{\ParticleCenter}$, 2D Covariance: $\bm{\Sigma'}$}
    \State $\lambda = \alpha^2 (3+\kappa) - 3$
    \State $\bm{x} = \text{SampleSigmaPoints}(\ParticleCenter, \bm{R}, \bm{S}, \lambda)$ \Comment{ \cref{eq:sigma_points}}
    \State $\bm{w} = \text{ComputeWeights}(\alpha, \beta, \lambda)$ \Comment{ \cref{eq:weights_1,eq:weights_2}}
    \State $\bm{v}_{\bm{x}} = \text{ProjectPoints}(\bm{x}, \bm{W}, \bm{D})$ \Comment { evaluate $g(\bm{x})$}
    \State $\bm{v}_{\ParticleCenter} = \text{EstimateMean}(\bm{v}_{\bm{x}}, \bm{w})$ \Comment{ \cref{eq:2d_conic_1}}  
    \State $\bm{\Sigma'} = \text{EstimateCovariance}(\bm{v}_{\ParticleCenter}, \bm{v}_{\bm{x}}, \bm{w})$ 
    \Comment{ \cref{eq:2d_conic_2}}
    \State \textbf{return} $\bm{v}_{\ParticleCenter}$ $\bm{\Sigma'}$
  \end{algorithmic}
\end{algo}
\end{minipage}

\section{Gaussian Rasterization Algorithm}
\label{sec:algorithm}
To show that our proposed UT-based projection can be used as a drop-in replacement to the 3DGS rasterization pipeline, we summarize their pipelines in terms of pseud-code in Algs.~\ref{proc:rasterize}~and~~\ref{proc:UT}. Note that we keep the Alg.~\ref{proc:rasterize} intact and only adapt the the \textsc{Estimate2DGaussian} function in Alg.~\ref{proc:rasterize}.

\section{Additional Experimental Results}

\begin{figure*}
    \centering
    \includegraphics[width=\linewidth]{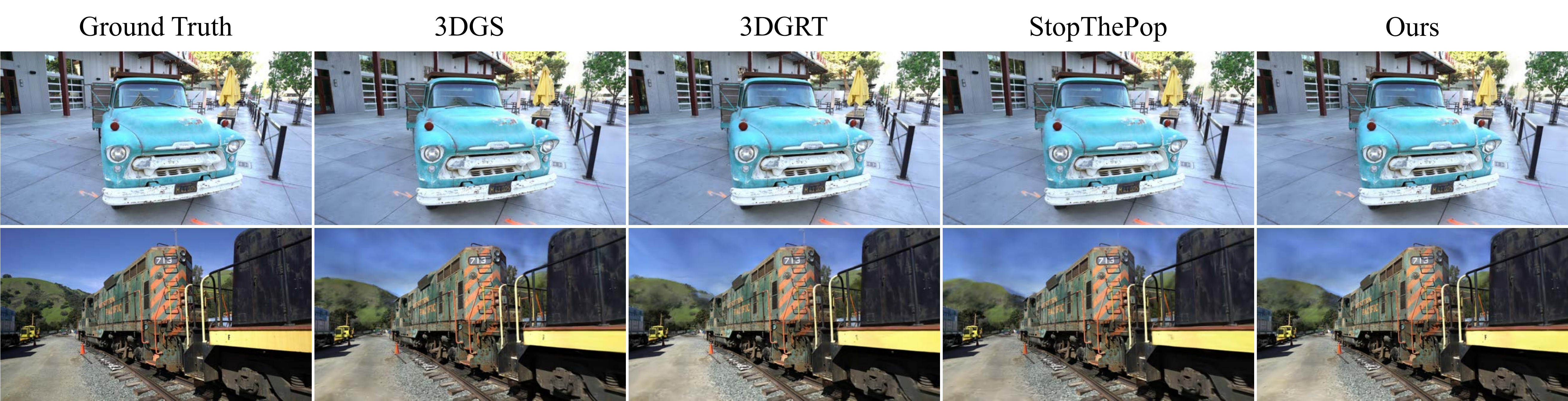}
    \caption{Qualitative comparison of our novel-view synthesis results against the baselines on the Tanks \& Temples~\cite{Knapitsch2017} dataset. }
    \label{fig:results_gallery_tandt}
\end{figure*}

In the main paper, \cref{fig:results_gallery} showcased a qualitative comparison of our model against various baselines on the MipNeRF360 dataset~\cite{barron2022mipnerf360}. Expanding on this, \cref{fig:results_gallery_tandt} provides an additional comparison using a different dataset (Tanks \& Temples~\cite{Knapitsch2017}). This figure highlights the qualitative performance of our method alongside the baseline approaches: 3DGS~\cite{kerbl3Dgaussians}, 3DGRT~\cite{3dgrt2024}, and StopThePop~\cite{radl2024stopthepop}. The results demonstrate that our approach delivers comparable or superior rendering quality.

\begin{table}
\caption{Detailed evaluation results of our methods on the Tanks \& Temples~\cite{Knapitsch2017} dataset.}
\vspace{-2mm}
\label{tab:per_scene_tandt}
\footnotesize
\centering
\begin{tabular}{ll|cccccc|cc}
  \toprule
  Method & Metric & \texttt{Train} & \texttt{Truck} \\
  \midrule
  \multirow{3}{*}{Ours} 
  & PSNR$\uparrow$ & 21.65 & 24.77 \\
  & SSIM$\uparrow$ & 0.813 & 0.868 \\
  & LPIPS$\downarrow$ & 0.199 & 0.157 \\
  \midrule
  \multirow{3}{*}{Ours (sorted)} 
  & PSNR$\uparrow$ & 21.39 & 24.41 \\
  & SSIM$\uparrow$ & 0.815 & 0.874 \\
  & LPIPS$\downarrow$ & 0.196 & 0.148 \\
  \bottomrule
\end{tabular}
\end{table}

\begin{table*}
\caption{Per-scene evaluation results of our methods on the MipNeRF360~\cite{barron2022mipnerf360} dataset}
\vspace{-2mm}
\label{tab:per_scene_mipnerf}
\footnotesize
\centering
\begin{tabular}{ll|ccccccccc}
  \toprule
  Method & Metric & \texttt{Bicycle} & \texttt{Bonsai} & \texttt{Counter} & \texttt{Garden} 
         & \texttt{Kitchen} & \texttt{Stump} & \texttt{Flowers} & \texttt{Room} & \texttt{Treehill} \\
  \midrule
  \multirow{3}{*}{Ours} 
  & PSNR$\uparrow$ & 
    24.21 & 32.17 & 29.03 & 26.90 & 31.23 & 26.51 & 21.48 & 31.64 & 22.15 \\
  & SSIM$\uparrow$ & 
    0.741 & 0.941 & 0.908 & 0.851 & 0.926 & 0.768 & 0.612 & 0.919 & 0.623 \\
  & LPIPS$\downarrow$ & 
    0.226 & 0.202 & 0.197 & 0.121 & 0.126 & 0.222 & 0.316 & 0.218 & 0.332 \\
  \midrule
  \multirow{3}{*}{Ours (sorted)} 
  & PSNR$\uparrow$ & 
    24.91 & 32.14 & 28.91 & 26.79 & 31.33 & 26.40 & 21.46 & 31.06 & 22.31 \\
  & SSIM$\uparrow$ & 
    0.756 & 0.940 & 0.907 & 0.851 & 0.926 & 0.768 & 0.610 & 0.919 & 0.629 \\
  & LPIPS$\downarrow$ & 
    0.217 & 0.200 & 0.195 & 0.121 & 0.124 & 0.223 & 0.318 & 0.213 & 0.323 \\
  \bottomrule
\end{tabular}
\end{table*}

\begin{table*}
\caption{Per-scene evaluation results of our methods on the Scannet++ dataset}
\vspace{-2mm}
\label{tab:per_scene_scannetpp}
\footnotesize
\centering
\begin{tabular}{ll|cccccc}
  \toprule
  Method & Metric & \texttt{0a5c013435} & \texttt{8d563fc2cc} & \texttt{bb87c292ad} & \texttt{d415cc449b} 
         & \texttt{e8ea9b4da8} & \texttt{fe1733741f} \\
  \midrule
  \multirow{3}{*}{Ours (sorted)}
  & PSNR$\uparrow$ & 
    29.80 & 27.09 & 31.28 & 27.75 & 33.09 & 25.59 \\
  & SSIM$\uparrow$ & 
    0.932 & 0.916 & 0.937 & 0.864 & 0.955 & 0.857 \\
  & LPIPS$\downarrow$ & 
    0.236 & 0.240 & 0.241 & 0.264 & 0.251 & 0.285 \\
  \bottomrule
\end{tabular}
\end{table*}

{
    \small
    \bibliographystyle{ieeenat_fullname}
    \bibliography{main}
}

\end{document}